\RequirePackage{ifpdf}
\ifpdf 
\documentclass[pdftex]{sigma}
\else
\documentclass{sigma}
\fi

\usepackage{mathrsfs,bm,ae,aecompl}

\newcommand{\ket}[1]{|#1\rangle}
\newcommand{\bra}[1]{\langle#1|}
\newcommand{\bracket}[2]{\langle#1|#2\rangle}
\newcommand{\wick}[1]{\bm:\!#1\!\bm:}
\renewcommand{\geq}{\geqslant}

\numberwithin{equation}{section}

\begin{document}

\allowdisplaybreaks

\renewcommand{\thefootnote}{$\star$}

\renewcommand{\PaperNumber}{073}

\FirstPageHeading

\ShortArticleName{Quantum Spacetime: a Disambiguation}

\ArticleName{Quantum Spacetime: a Disambiguation\footnote{This paper is a
contribution to the Special Issue ``Noncommutative Spaces and Fields''. The
full collection is available at
\href{http://www.emis.de/journals/SIGMA/noncommutative.html}{http://www.emis.de/journals/SIGMA/noncommutative.html}}}

\Author{Gherardo PIACITELLI}

\AuthorNameForHeading{G.~Piacitelli}

\Address{SISSA, Via Bonomea 265, 34136, Trieste, Italy}
\Email{\href{mailto:gherardo@piacitelli.org}{gherardo@piacitelli.org}}

\ArticleDates{Received April 29, 2010, in f\/inal form September 15, 2010;  Published online September 23, 2010}

\Abstract{We review an approach to
non-commutative geometry, where models are constructed by quantisation
of the coordinates. In particular we focus on the full DFR model
and its irreducible components; the (arbitrary) restriction
to a particular irreducible component is often referred to
as the ``canonical quantum spacetime''.
The aim is to distinguish and compare the approaches under various
points of view, including motivations, prescriptions for quantisation,
the choice of mathematical objects and concepts, approaches to dynamics
and to covariance.}


\Classification{46L65; 81T75; 83C65; 58B32}

{\small \tableofcontents}

\renewcommand{\thefootnote}{\arabic{footnote}}
\setcounter{footnote}{0}

\section{Introduction}

Within the class of models of quantum spacetime which are def\/ined
by imposing non trivial commutation relations on the coordinates,
there are quite dif\/ferent approaches on the market.
A list of dif\/ferences should include at least
\begin{itemize}\itemsep=0pt
\item[a)]  dif\/ferent aims and motivations,
\item[b)]  dif\/ferent philosophies in the quantisation prescriptions,
\item[c)]  dif\/ferent classes of algebras (as mathematical objects),
\item[d)] dif\/ferent approaches to dynamics and,
\item[e)]  last but not least, dif\/ferent approaches to covariance.
\end{itemize}
Of course the above dif\/ferences are not all independent.

In this review, some comments on the above issues will be made under
a narrow perspective. I will not attempt a thorough description
of the immense literature, both for reasons of space and lack of knowledge;
I apologise in advance for any unforgivable omission. Very little of the
material and the comments presented is new, although some remarks
only previously arose in discussions.

The root of motivations for these investigations
is shared by everybody in this f\/ield:
the hope that a full theory of quantum
gravity could possibly emerge from completing the transition to a full
quantum ``description of nature'',
encompassing geometry as well.
Another common key remark is that the ultraviolet nightmare
which plagues perturbative quantum f\/ield theory might be a
symptom that geometry is not classical in the small. Moreover,
by a consistence argument, sharp localisation should
not be expected to be possible at all scales: indeed, at very small scales
it should lead to some instability of the geometric background, as
localisation in sharp regions might induce the formation of closed
horizons. This comment is very old and took many shapes; depending on the
chosen shape, it leads to dif\/ferent approaches to the problem; we will
discuss it in Section~\ref{sec:motivations}. There, we also will shortly comment
on Doubly Special Relativity, and show that singly Special Relativity is
already multiply special, at least in some models.

On a more detailed ground, the above
comments have been interpreted
in at least three dif\/ferent contexts: string theory, quantum
f\/ield theory, and quantum mechanics.

We will not discuss the
string-theoretical aspects
because I feel incompetent.

In quantum f\/ield theory,
the quantisation of the
localisation algebra should be interpreted
as the framework providing
a non-commutative replacement for pointwise products of local free
f\/ields in interaction terms; we will discuss this in Section~\ref{sec:qft_qst}.

\looseness=-1
The quantum mechanical approach, instead, involves
a change in the commutation relations
for the momenta as well: while in a non-commutative algebra
(with translation covariance)
momenta pairwise commute and generate the usual action of the classical
translation group, in this approach\footnote{\label{fn:NC}These approaches
are usually referred to as
``non-commutative quantum f\/ield theory (NQFT)'' and
``non-commutative quantum mechanics (NCQM)''.
However, since by its very nature quantum physics is non
commutative, ``non-commutative quantum'' contrasted with ``quantum'' alone
is really an awful terminology, in my humble opinion. Since words do matter,
we shall dismiss it here.}
momenta and coordinates with mixed commutation relations should ideally
replace the usual
Schr\"odinger operators at small scales. We will not discuss this approach;
indeed, because of the extremely high energies expected to
occur in processes at Planck scale, a Planck scale modif\/ication of quantum
mechanics  does not seem to be a physically interesting limit.

The choice of the relevant algebraic structure and
quantisation prescription is somewhat critical,
and is full of consequences on the technical and conceptual
ground as well. The issue is not at all a marginal technical one.
For example, it is often said that to a great extent quantum physics
is a spectral theory. But if the relations
\[
[\bm x^\mu,\bm x^\nu]=i\theta^{\mu\nu}\bm I,
\]
instead of being understood as regular relations among selfadjoint operators
on some Hilbert space,
are taken as the def\/ining relations between a f\/inite set
of Hermitean generators of an abstract
\mbox{$*$-al}ge\-bra (which turns out not $C^*$), then every non trivial
element in that algebra (including the
selfadjoint coordinates \(\bm x^\mu\))
has the whole complex plane as its spectrum, and the spectral theory
is completely trivial.  Moreover, the transition from the above relations
to the apparently equivalent notation
\begin{equation}
\label{eq:not_innocent}
\bm x^\mu\star\bm x^\nu-\bm x^\nu\star\bm x^\mu=i\theta^{\mu\nu}
\end{equation}
is far from innocent:
if \(\star\) is understood as a twisted product of symbols,
the above cannot coexist with the so called Weyl quantisation: they are
mutually exclusive. This will be discussed in Section~\ref{sec:which_algebra}.

The importance of identifying the representations of the relations is
made even more
evident when we consider for example time/space commutativity; we will
see in Section~\ref{subsec:sad_fate} that there is a variety of situations,
including the possibility that no representation exists.

The full DFR model is described in some detail in Section~\ref{sec:dfr}, where an attempt is made to make some
concepts more accessible, notwithstanding the technical implications.
The material is organised so to easily discuss related concepts
available in the literature.

In Section~\ref{sec:CQST} we comment on some f\/lavours
of the ``canonical'' quantum spacetime. In particu\-lar the
relations with twisted covariance
are discussed in Section~\ref{subsec:tw_cov}.

Section~\ref{sec:qft_qst} is devoted to quantum f\/ield theory (QFT) on
quantum spacetime (QST). We will recall the approach based on the
Gell-Mann--Low formula with a non local interaction, and focus on the
issue of time-ordering, and its relations with unitarity. We also shortly
recall some basic facts about the diagrams, with a disambiguation between the
Filk rules and those arising from the Dyson series. We f\/inally will shortly
comment on the availability of Euclidean methods, and on the IR/UV mixing.

We will not conclude with an outlook, for which I refer instead to~\cite{Doplicher:2001qt,Doplicher:2006vy}.

\section{Motivations}
\label{sec:motivations}
\subsection{Planck length and probes}
If spacetime is classical (e.g.\ the Minkowski spacetime), the non trivial
Heisenberg space/mo\-men\-tum uncertainty relations
\[
\Delta x_j\Delta p_j\geq \frac\hbar 2
\]
should be ideally complemented with the missing \(4^{\text{th}}\) component
time/energy uncertainty relation
\[
\Delta t\Delta E\geq \frac\hbar 2
\]
(all other uncertainty relations being trivial).
Indeed in classical Hamiltonian mechanics, time and energy are canonically
conjugate variables.

Although there are no time and position {\itshape
observables} in a relativistic theory, the above relations can be taken as an
indication that the localisation process implies an energy transfer to
spacetime. The higher is the precision of localisation in time
(small \(\Delta t\)), the higher
is the energy transferred to the geometric background; if localisation is also
conf\/ined
\begin{equation}\label{eq:text_small_volume}
\text{in a small space volume,}
\end{equation}
then the energy density induced by localisation might produce a closed
horizon (a black hole). This would lead to a paradoxical situation, where
the horizon would
trap any information, and we would face a localisation process with no output.
Thus, in order to have a consistent {\itshape operational}
description of spacetime, geometry should be modif\/ied in the small so
to prevent ``too sharp'' localisation.

This remark is very old and it is probably fair to ascribe it to folklore.
As far as I know, the f\/irst who tried to give it a quantitative content was
Mead in \cite{Mead}\footnote{Actually that paper was
submitted in 1961, but underwent referee troubles; see the interesting letter
of Mead to Physics Today~\cite{Mead_letter}.}. From a quite general argument
based on the ``Heisenberg microscope'',
he deduced that, to avoid the above mentioned spacetime instability, the
size \(\Delta r\) of probes should be bounded below by the
Planck length
\[
\lambda_P=\left({G\hbar\over
c^3}\right)^{1/2}\simeq1.6\times10^{-33}~\text{cm.}
\]
Reduced to its essence, Mead's argument is based on the remark that
if one considers a mass~\(m\) with its associated Compton wavelength
\(\lambda(m)\) and Schwarzschild radius \(R(m)\), the condition
\(\lambda(m)\sim R(m)\) is fulf\/illed for \(m\sim m_P\), the Planck mass,
and we f\/ind \(\lambda(m_P)\sim R(m_P)\sim\lambda_P\).

Mead deduced from this the necessity
of a minimal uncertainty of order of \(\lambda_P\).
While this statement is somewhat questionable,
certainly the basic argument provides a good motivation for assuming
that the Planck length is the relevant  scale where gravitation and quantum physics should meet.

Analogous conclusions were drawn by Amati, Ciafaloni and Veneziano, from
a dif\/ferent perspective \cite{Amati:1987wq} and in the context of string
theory; they found a relation of the form
\begin{equation}\label{eq:Mead_rel}
\Delta r\geq \frac{\hbar}{\Delta p}+\alpha'\Delta p,
\end{equation}
where \(\alpha'\) is a positive constant depending on Regge's slope and the
gravitation constant.
For positive \(x\)'s, \(\hbar/x+\alpha' x\)
takes its minimum value \(\lambda=2\sqrt{\alpha'\hbar}\) at
\(x=\sqrt{\hbar/\alpha'}\). Hence they found again the lower estimate
\[
\Delta r\geq \lambda_P\sim 10^{-33}~{\rm cm}
\]
for the size of an admissible probe.

The same relation \eqref{eq:Mead_rel}
and the same conclusions were reobtained by Maggiore
\cite{Maggiore:1993rv}, with a~``model independent'' argument quite close
to that of Mead; probably Maggiore was not aware of Mead's analysis.

\subsection{Planck length and uncertainty relations}
The argument of Mead and Maggiore is indeed model independent
from the dynamical point of view, but there is a hidden
assumption \cite{Doplicher:2006vy},
which also is present in \cite{Amati:1987wq}.
Namely that the uncertainties of the three space coordinates must be of the
same order.
Indeed, to derive \eqref{eq:Mead_rel} the interaction of probes with spacetime
is explicitly modeled (at least by Mead and Maggiore) by means of a solution
of the Einstein equations with spherical symmetry: the ``\(\Delta (p)\)''
showing up in \eqref{eq:Mead_rel} is the total black hole momentum.
Hence the only conclusion
which is obtained in those references is that
\begin{quote}
{\itshape
If we assume that spacetime only can be probed with spherically symmetric
devices, than there must be a minimal uncertainty.}
\end{quote}

But why should probes be spherical?
Let us come back to the heuristic discussion at the beginning of the section.
All the point is hidden
in \eqref{eq:text_small_volume}. Indeed, if we assume that
the localisation region has spherical symmetry, then all the \(\Delta x_j\)'s
are equal to each other, and in particular they all
must be small in order to obtain a small volume. But
 a small volume can also be obtained with some
large \(\Delta x_j\); it only is necessary that \(\prod_j\Delta x_j\)
is small. Hence, if we dismiss the assumption of spherically symmetric
probes, the argument of Mead
for the necessity of a minimal uncertainty
for a length measurement is not compelling
any more\footnote{The situation is much alike that of usual quantum
mechanics (for a particle
on the line, for simplicity). There the Heisenberg uncertainty relations
\(\Delta \bm P\Delta \bm Q\geq \hbar/2\) do not imply that $\bm P$, $\bm Q$
have discrete  spectrum. On the contrary, it is well known that
the Schr\"odinger operators have purely continuous spectrum. The rectangular
Planck cells used in quantum statistical mechanics have no f\/ixed shape;
for all computations it is suf\/f\/icient to know their
area \(\hbar/2\).}. Indeed, this remark was the starting point of the
investigation of~\cite{Doplicher:1994tu}.

It is interesting to note that Maggiore was able to rederive the relation
\eqref{eq:Mead_rel} from suitably
modif\/ied commutation relations between positions and momenta
\cite{Maggiore:1993kv}. He found that the
so called \(\kappa\)-Poincar\'e commutation
relations induce uncertainty relations such that
\begin{equation}\label{eq:text_regime}
\text{in a certain regime}
\end{equation}
the relation \eqref{eq:Mead_rel} is fulf\/illed. However, the close
\eqref{eq:text_regime} implies that those relations are not always valid;
they only are asymptotically valid under limiting conditions which
imply spherical symmetry.
Maggiore then concluded, once again, that there must be a minimal uncertainty;
but if one looks instead
at the {\itshape unrestricted} uncertainty relations,
no absolute bounds can be deduced for the values of a single
coordinate.

Starting from a quantitative heuristic analysis (using quantum f\/ield theory)
of the initial remark of this section, Doplicher, Fredenhagen and Roberts
wrote down operationally motivated uncertainty relations:
\begin{gather*}
\Delta x^0(\Delta x^1+\Delta x^2+\Delta x^3)\geqslant \lambda_P^2,\\
\Delta x^1\Delta x^2+\Delta x^1\Delta x^3+
\Delta x^2\Delta x^3
\geqslant \lambda_P^2.
\end{gather*}
Moreover, they found commutation relations inducing those uncertainty
relations \cite{Doplicher:1994tu}. This
led them to formulate their model (DFR model); we shall discuss
those relations in some detail, later in this review. Let us anticipate however
that no minimal length for a length measurement shows up in that model:
coordinates have purely continuous spectrum  and nothing prevents sharp
localisation in some of the coordinates; it is only the precision in the
simultaneous localisation of all the coordinates which is bounded
by uncertainty relations involving the Planck length as a characteristic length
of the model.

\subsection{Spacetime regularisation}

We may observe, as an additional motivation for spacetime quantisation,
that relativistic
quantum f\/ield theory has defeated for almost a century every attempt to
construct a non perturbative model in physical (i.e.\ 4) dimensions: only
free f\/ields are known as exact models. Moreover, the perturbative approach
is plagued by severe divergences, called ``ultraviolet'' because they involve
the behaviour at small distances. Notwithstanding the general success of
renormalisation theo\-ry\footnote{A rather complicate recursive local covariant
strategy for removing divergences from the perturbative series.}
and the wonderful experimental validation of lower
order renormalised contributions, no exact perturbative limit theory is
known in \(d=4\),
which is  not trivial (i.e.\ not free). It is rather reasonable to conclude that
some internal contradiction in the theory arises when we push locality
to the ``inf\/initely small scale''. Hence an additional motivation for spacetime
quantisation is the hope for obtaining well def\/ined interacting models, as a
byproduct of the regularising ef\/fect of the quantum texture of spacetime.

For the sake of completeness, we mention that the f\/irst model of quantum
spacetime based on non-commuting coordinates is due to Snyder
\cite{Snyder:1946qz}.  This model was
motivated precisely as an attempt to mimic lattice regularisation of the
perturbative terms
in a covariant way. The elegant idea was abandoned when the renormalisation
program gave an ef\/fective solution to the perturbative problem.
Although Snyder was aware of the conceptual implications of spacetime
quantisation and of the possible follow-up of his research, he did not develop
this line. Indeed, from his point of view uncertainty relations were a
possible source of nuisance, and he only computed them to check that they
would have not spoiled the consistency of the picture underlying his framework
(see especially the very last sentence of his paper).

\subsection{The meaning of ``minimal length'', and Doubly Special Relativity}

The notation ``\(\Delta x^\mu\)'' deserves to be better specif\/ied. It is of
course related with the precision with which the position
\(x^\mu\) can be known. It is however necessary to distinguish the concepts
of minimal length and minimal uncertainty.

If the position is described by a selfadjoint operator
\(\bm x^\mu\) under a quantum perspective, then it is understood that the
uncertainty of the position in a state \(\omega\) is given by
\[
\Delta x^\mu=\Delta_\omega(\bm x^\mu)=
\sqrt{\omega(({\bm x^\mu}-\omega(\bm x^\mu)\bm I)^2)}=
\sqrt{\omega({\bm x^\mu}^2)-\omega(\bm x^\mu)^2}.
\]
It is clear that, for any spectral value
\(a\) of \(\bm x^\mu\), there is a
state \(\omega\) such that  \(\omega(\bm x^\mu)=a\) with arbitrarily small
uncertainty\footnote{If \(a\) is an eigenvalue with eigenvector
\(\ket a\), take \(\omega(a)=\bra a\cdot\ket a/\bracket aa\);
if \(a\) is in the continuous spectrum, replace the generalised eigenket
\(\ket a\) with a suf\/f\/iciently good f\/inite length approximation.},
whatever commutation relations the coordinates do fulf\/il. Hence an absolute
lower bound for the uncertainty of one coordinate is incompatible with the
possibility of describing the coordinates by means of operators.

For example, since Snyder's model is def\/ined in terms of selfadjoint
coordinates, it is always possible to sharply localise in one Snyder
coordinate alone.
There, however, interest is appointed to the fact that
the spectrum of each coordinate \(\bm x^\mu\) is of the
form \(\lambda\mathbb Z\) for some parameter \(\lambda\). In this model we may
speak of a minimal length, def\/ined
as the minimal separation \(\Delta x^\mu\) between
two possible values of the same coordinate (which plays the r\^ole of the precision of an instrument, instead of its error); but this now means to attach a
dif\/ferent meaning to the notation \(\Delta x^\mu\).
Anyhow, we see here a f\/irst example
where the concept of minimal length has a meaning even when all the coordinates
can be {\itshape separately} sharply localised.

In general, relative bounds on the precision of joint localisation in two or
more non commuting coordinates arise,
as a consequence of the commutation rules, by
(variants of) the generalised Heisenberg uncertainty theorem\footnote{For any
$\bm A=\bm A^*$, $\bm B=\bm B^*$ and state \(\omega\),
\(\Delta_\omega(\bm A)\Delta_\omega(\bm B)\geqslant
\frac12|\omega([\bm A,\bm B])|\).}. In the case of Snyder's model,
such uncertainty relations are not known (Snyder never published the
computations he made). In the case of the DFR model such relations are known,
and actually were the motivations for the investigation.

In the DFR model, though the coordinates are selfadjoint and have spectrum
\(\mathbb R\), there also is room for a minimal length. Indeed the
``Euclidean square distance''\footnote{From the (classical) origin; but note
that the model has translation covariance.}
operator is bounded below by \(2\lambda_P^2\).
Note that the Euclidean length is not a Lorentz invariant, so that inequivalent
observers may fail to agree on states achieving the minimum. However,
the minimal length (in the sense above) is well def\/ined and has the same
value  for every observer, and is thus
a general property of the model.

Above, we speak of Euclidean distance of a quantum event from a classical
point, which has no physical interpretation. However there is a more
ref\/ined version of this  comment, which shows that there also is a minimal
Euclidean distance between two independent quantum events~\cite{Bahns:2003vb}.

It is often stated that the existence of a minimal\footnote{The ``smallness''
of this length is not relevant here; we'd better call it a universal length.}
length is incompatible
with ordinary Lorentz covariance, because of Lorentz--Fitzgerald contraction;
and that, as a consequence, it is necessary to ``deform'' covariance in some
appropriate sense (e.g.\ in the sense of Quantum Groups).
This in turn is interpreted in the general phenomenological framework
of ``Doubly Special Relativity'' (DSR). We will
not enter here in a description
of DSR (see e.g.~\cite{AmelinoCamelia:2010pd}).
We only observe that the DFR analysis
provides a well def\/ined {\itshape model}
where
\begin{itemize}\itemsep=0pt
\item the coordinates are described by non commuting coordinates, which are selfadjoint operators on some Hilbert space,
\item there is a continuous unitary representation of the
({\itshape usual, undeformed})
Poincar\'e group on the same Hilbert space,
\item the above representation implements {\itshape ordinary}
covariance of the coordinates
(quantum symmetry in the usual sense of Wigner), and
\item there is a minimal length {\itshape which is the same for all equivalent observers}.
\end{itemize}
Hence Lorentz--Fitzgerald contraction is not at all incompatible with the
presence of a minimal length; the quest for DSR
does not forces us in principle to deform
covariance, though of course deformed covariance might well be anyway
interesting for  other reasons.

In a sense, usual special relativity is already multiply special. To encompass
\(n\)-ply special relativity it is suf\/f\/icient to f\/ind
non commutative algebras with \(n\) universal parameters, equipped with an
undeformed  action
of the usual Poincar\'e group by automorphisms; there is plenty of such
examples (see e.g.~\cite{Dabrowski:2009mw}).

\section{Which algebra?}
\label{sec:which_algebra}

In this section we will begin our disambiguation by
discussing and comparing some necessary mathematical tools and concepts.
In particular we will show that def\/ining the algebra of quantum spacetime
through Weyl quantisation, or as an algebra generated by f\/initely many
elements and relations,
leads to profoundly dif\/ferent mathematical structures, and we will shortly
discuss the consequences of this.

Let us keep apart for the moment the issue of covariance, and consider
relations of the form
\begin{equation}\label{eq:rels}
[\bm x^\mu,\bm x^\nu]=i\theta^{\mu\nu}\bm I,
\end{equation}
where \(\theta^{\mu\nu}\) is a non zero
antisymmetric matrix with real entries. Assume
moreover that the \(\bm x^\mu\) are selfadjoint.

Let us show that a $*$-algebra \(\mathcal A_\theta\) with unit, which contains
selfadjoint elements \(\bm x^\mu\) fulf\/illing the above relations, cannot
be faithfully represented as an algebra of bounded operators.

To see this\footnote{For a nice purely $C^*$-algebraic argument not relying on
von Neumann uniqueness, see \cite[\S~11]{cuntz-weyl-algebra}},
let $\bar\mu$, $\bar\nu$ be some f\/ixed choice of the values of the dummy
indices $\mu$, $\nu$, such that \(\theta^{\bar\mu\bar\nu}\neq 0\). If we rename
$\bm P=\bm x^{\bar\nu}$, $\bm Q=\bm x^{\bar\mu}$, $\hbar=\theta^{\bar\mu\bar\nu}$,
we discover the well known Canonical Commutation Relation (CCR)
\[
[\bm P,\bm Q]=-i\hbar \bm I
\]
hidden in the spacetime commutation relations (as a purely mathematical fact,
deprived of any physical interpretation).
It is well known from elementary quantum mechanics that the CCR cannot be
represented by means of bounded operators: at least one among position
and momentum must be unbounded. It follows that the algebra
\(\mathcal A_\theta\) cannot have a faithful $*$-representation by bounded
operators on some Hilbert space (for, otherwise, there also would be a
bounded representation of the CCR). This applies in particular if \(\mathcal
A_\theta\) is precisely the algebra generated by the relations
\eqref{eq:rels},
as we shall assume from now on.

In other words,
\(\mathcal A_\theta\) has no $C^*$-completions at all (technically: there is no
pre-$C^*$-norm on~\(\mathcal A_\theta\)).

Existence of $C^*$-completions is not a marginal technical detail for quantum
physics. In a~generic
$*$-algebra, all one can say about the spectrum of a selfadjoint element is
that it is invariant under complex
conjugation; but it might well fail to be contained by the real axis. Moreover,
``functions'' of elements in a generic $*$-algebra (even a Banach one)
may be def\/ined to a limited
extent. The $C^*$ property\footnote{A $C^*$-algebra is a Banach $*$-algebra enjoying
\(\|a^*a\|=\|a\|^2\) for any of its elements \(a\). A $C^*$-algebra admits
a~unique $C^*$-norm, fully determined by the algebraic structure.}
is the minimal requirement ensuring that spectra
of selfadjoints are real, and that there is a useful notion of functional
calculus, mapping the spectra in the natural way. These two properties,
in turn, are indispensable for a sound physical interpretation.

The existence of suf\/f\/iciently many
representations of (regular) commutation
relations is a~{\itshape conditio sine qua non}
for the existence of $C^*$-completions; whence comes the quantum motto
\begin{center}
{\scshape ``No deformation without representation!''}
\end{center}

A completely dif\/ferent approach to the construction of the spacetime
algebra was proposed much earlier in \cite{Doplicher:1994tu}, in a more general setting.
The idea
is to follow the approach proposed by Weyl (and adopted by von Neumann)
in the case of quantum mechanics,
where a canonical map is def\/ined, which sends functions
into operators, and is called Weyl quantisation; the functions in the domain of the quantisation are called ``symbols''. By pulling back the operator product
to symbols, one obtains the
$*$-algebra \(\mathcal E_\theta\) of ``symbols'', equipped with
a non-commutative product \(\star\) which is a deformation of the usual
pointwise product (e.g.\ in the sense of~\cite{rieffel-lincei});
we will describe it in more detail here below. It was shown
in \cite{Doplicher:1994tu} that, for a large class of matrices \(\theta\), this
$*$-algebra  may be faithfully represented as a $*$-algebra of
bounded operators on some Hilbert space ($=$~it has a unique $C^*$-completion).
Note however that, though this $*$-algebra
is indirectly def\/ined by commutation relations among the coordinates,
{\itshape it does not contain the coordinates themselves!} The quantum
coordinates \(\bm x^\mu\) arise as the {\itshape unbounded} selfadjoint
operators on the Hilbert space which are canonically
associated with the Weyl
quantisation.

Hence, the $*$-algebras
$\mathcal A_\theta$, $\mathcal E_\theta$ are non-isomorphic in a very substantial
way, which can not be circumvented, and which leads to very dif\/ferent
frameworks, both on the conceptual and technical side.

As a remarkable example, note
that in the algebra \(\mathcal A_\theta\) the Weyl quantisation is
not available, since there is no such object in \(\mathcal A_\theta\) as
\(e^{ik_\mu \bm x^\mu}\); nor there is any natural notion of completion
or multipliers algebra of \(\mathcal A_\theta\), where
to give a meaning to the def\/ining series of \(e^{ik_\mu \bm x^\mu}\). Even worse
is that the spectral theory is completely trivial in \(\mathcal A_\theta\),
as we shall soon see.

In the remaining of this section we will describe more precisely the way
in which the algebras~\(\mathcal E_\theta\) and~\(\mathcal A_\theta\)
are constructed, and we will make some of the above remarks
more precise. We also will comment in some detail on
the most dangerous cowboy downtown,
the Moyal expansion. Since in the literature the same words and symbols
are used in
the two dif\/ferent contexts for related concepts,
we will be a little pedantic, in
order to detect the sources of ambiguity. Although references will be made
to advanced mathematical concepts at some points,
hopefully the bulk material should appear reasonably accessible.

\subsection{Weyl relations and quantisation}
\label{subsec:weyl_rels}
By the BCH formula, \eqref{eq:rels} formally implies the relations
\begin{equation}\label{eq:weyl_rels}
e^{ik_\mu\bm x^\mu}e^{ih_\nu\bm x^\nu}=
e^{-\frac i2h_\mu\theta^{\mu\nu}k_\nu}e^{i(h+k)_\mu\bm x^{\mu}}.
\end{equation}
Contrary to \eqref{eq:rels}, \eqref{eq:weyl_rels} is completely
unambiguous if the \(\bm x^\mu\) are understood to be selfadjoint operators.

By a simple argument (\cite{Doplicher:1994tu}; see also
Section~\ref{subsec:exist_coord})
it is possible to show that, for a large class of matrices \(\theta\),
there exist unbounded selfadjoint
operators \(\bm x^\mu\) on a Hilbert space \(\mathfrak H\), fulf\/illing the
above relations as relations between unitary operators. For a smaller,
yet still large class of matrices, the operators \(\bm x^\mu\) are also
unique (up to multiplicity and equivalence); we assume this to be the case.

In \cite{Doplicher:1994tu}
it also was proposed to adapt the Weyl quantisation prescription
\begin{equation}\label{eq:weyl_q}
f(\bm x)=\int dk\;\check f(k)e^{ik_\mu \bm x^\mu},
\end{equation}
where
\[
\check f(k)=\frac1{(2\pi)^4}\int dx\;f(x)e^{-ik_\mu x^\mu}
\]
is the usual Fourier transform. The idea is to replace the plane
waves with their quantised counterparts. Note that
\(\bar f(\bm x)=f(\bm x)^*\), where \(\bar f\) is the function obtained by
pointwise complex conjugation, and \({}^*\) means adjoint (as an operator).
In particular \(f(\bm x)\) is selfadjoint if \(f\) is real.

Note that, in order that both integrals
are well def\/ined, we require that both $f$, $\check f$ are \(L^1\);
such functions are called symbols; in particular the coordinate
functions \(x^\mu\) are {\itshape not} in the domain of the quantisation.

Since the map \(f\mapsto f(\bm x)\) is injective,
it is possible to def\/ine an associative product of symbols by
setting
\begin{equation}
(f\star g)(\bm x)=f(\bm x)g(\bm x),
\end{equation}
where on the right hand side the operator product is taken.
Standard computations using \eqref{eq:weyl_rels} and antisymmetry of \(\theta\)
yield
\begin{equation}
f(\bm x)g(\bm x)=(f\star g)(\bm x)=\widehat{f\tilde\times g}(\bm x),
\end{equation}
where
\begin{equation}\label{eq:tw_conv}
(\check f\tilde\times\check g)(k)=\int dh\;\check f(h)\check g(k-h)e^{-\frac i2 h_\mu\theta^{\mu\nu}k_\nu}.
\end{equation}
More explicitly,{\samepage
\begin{equation}\label{eq:tw_prod_expl}
(f\star g)(x)=\int dk\;e^{ik_\mu x^\mu}
\int dh\;\check f(h)\check g(k-h)e^{-\frac i2 h_\mu\theta^{\mu\nu}k_\nu}.
\end{equation}
The product \(\star\) is called the twisted product; \(\tilde\times\) is the
twisted convolution.}

The algebra \(\mathcal E_\theta\) of symbols equipped with the product \(\star\)
and the involution \(f\mapsto \bar f\) is a well def\/ined, associative
$*$-algebra. Note that only the relations \eqref{eq:weyl_rels}
were used to def\/ine it;
at this stage it is not necessary to know the
precise form of the operators \(\bm x^\mu\). However, knowing that they exist
and are unique is suf\/f\/icient to conclude that this $*$-algebra admits a
unique $C^*$-completion, which is found to be the $C^*$-algebra \(\mathcal K\)
of compact operators, see \cite{Doplicher:1994tu}.

Note that the $C^*$-completion of \(\mathcal E_\theta\) is the same
for all \(\theta\)'s as above; hence the particular choice of \(\theta\)
is visible only in the quantisation prescription \eqref{eq:weyl_q},
while the $C^*$-algebra
is the same for all choices.

Finally, observe that equation \eqref{eq:not_innocent} is meaningless
in this framework.
For unbounded operators, even the operator
product is  a concept of little interest in general, because of problems with
domains (operator product is composition of maps). Hence the notation
\(\bm x^\mu\star \bm x^\nu\) can not even be given a sound meaning as a product
of operators. Only the Weyl relations
\eqref{eq:weyl_rels} are unambiguously meaningful;
in this context \eqref{eq:rels}
is always understood as a symbolic reminder of the Weyl relations.
\subsection{Moyal expansion}
\label{subsec:moy_exp}
Let $f$, $g$ be symbols such that $\check f$, $\check g$ have compact support.
Note that by the Paley--Wiener theorem this implies that $f$, $g$ are entire
analytic functions of \(\mathbb R^4\). It is important to keep in mind that
they also are symbols, thus \(L^1\); in particular, polynomials are ruled
out.

We may now replace the exponential \(\exp(-i\theta^{\mu\nu}h_\mu k_\nu)\) by its series expansion; for the selected subclass of symbols,
\(\int\sum=\sum\int\), and
\[
(\check f\times \check g)=
\sum_{n=0}^\infty\frac {(-i/2)^n}{n!}\int dk\;
(h_\mu\theta^{\mu\nu}k_\nu)^n\check f(k)\check g(k).
\]
By standard Fourier theory
\[
(f\star g)(x)=\widehat{\check f\times \check g}(x)=
\sum_{n=0}^\infty\frac {(i/2)^n}{n!}
m\left((\theta^{\mu\nu}\partial_\mu\otimes\partial_\nu)^nf\otimes g\right)(x),
\]
where \(m(f\otimes g)(x)=f(x)g(x)\). As a shorthand of the above
(Moyal) expansion, one writes
\[
f\star g=m\circ e^{\frac i2\theta^{\mu\nu}\partial_\mu\otimes\partial_\nu}
f\otimes g,
\]
which is unambiguously true for real analytic symbols.

The Moyal expansion is often said asymptotic. However this refers to the
asymptotic behaviour of the truncated series when \(\theta\) goes to zero;
see \cite{Estrada:1989da}. There is an extensive literature
about extensions of the Moyal expansion
to wider classes of functions than real-analytic symbols.
However, the underlying
philosophy there~--~as far as I know~--~is to focus the
attention on more general, standalone non local
products, without correspondingly enlarging the domain of the
quantisation. For this reason we will not discuss them in this review.

\subsection{Algebraic relations}

Usually the $*$-algebra generated by a f\/inite set of elements
with given commutation relations and adjoints is def\/ined as the
free algebra generated by that set, divided by the equivalence relation
induced by the given commutation relations. However in our case there is a
simpler way.

Consider the ring \(\mathbb C[\bm x^\mu]\) of polynomials in the variables
$\bm x^0$, $\bm x^1$, $\bm x^2$, $\bm x^3$. With
\(fg\) the usual product of two polynomials $f$, $g$, def\/ine the map
\[
m: \ f\otimes g\mapsto fg.
\]
The expression
\[
\mathcal F_\theta=e^{\frac i2\theta^{\mu\nu}\partial_\mu\otimes\partial_\nu}
\]
gives a well def\/ined linear operator on
\(\mathbb C[x^\mu]\otimes\mathbb C[x^\mu]\), since for any two
polynomials \(f,g\in \mathbb C[x^\mu]\),
only a f\/inite number of terms in the sum
\[
\mathcal F_\theta f\otimes g=
\sum_{n=0}^\infty\frac {(i/2)^n}{n!}
(\theta^{\mu\nu}\partial_\mu\otimes\partial_\nu)^nf\otimes g
\]
is dif\/ferent from 0.

We now may set
\[
f\star g=m\circ F_\theta f\otimes g.
\]
One may directly check that the above def\/ines an associative product of
polynomials. Moreover, due to the antisymmetry of \(\theta\), it is also
involutive under the usual involution of polynomials (complex conjugation
of coef\/f\/icients). The unital $*$-algebra \(\mathcal A_\theta\)
obtained by equipping \(\mathbb C[x^\mu]\)
with the usual involution and the product \(\star\) is generated by
its elements \(\bm x^\mu\), which are selfadjoint and fulf\/il
\[
\bm x^\mu\star \bm x^\nu-\bm x^\nu\star\bm x^\mu=i\theta^{\mu\nu}.
\]
Since the algebra generated by a f\/inite set of
relations is unique (up to isomorphisms), here we are.

Notations: \(\bm x^{\mu_1}\dotsm \bm x^{\mu_k}\) is the product taken
in \(\mathbb C[\bm x^\mu]\); while
\(\bm x^{\mu_1}\star\dotsm\star\bm x^{\mu_k}\) is taken in \(\mathcal A_\theta\).
Note in particular that, due to antisymmetry of \(\theta\),
\((\bm x^\mu)^n=(\bm x^\mu)^{\star n}\); moreover,
\(\bm x^\mu\star \bm x^\nu=\bm x^\mu\bm x^\nu+(i/2)\theta^{\mu\nu}\).

This algebra contains in particular the elements \(\bm x^\nu\), and thus
admits no $C^*$-completion, as discussed in the f\/irst part of this section.

It is easily seen that the (maximal)
degree of polynomials is additive under the deformed
product \(\star\) (the non-commutative corrections \(f\star g-fg\) being
of lower degree). Hence, like in \(\mathbb C[\bm x^\mu]\),
no polynomial of non zero degree can be
invertible in \(\mathcal A_\theta\).
It follows that any element on nonzero degree has the whole
complex plane as its spectrum (because resolvents never exist),
no matter whether it is selfadjoint,
skewadjoint, or none of the two. Moreover, since the only invertible elements
are the multiples of the identity, the only unitary elements in \(\mathcal A_\theta\) are the complex phases.

Note that in particular
\[
\text{spec}_{\mathcal A_\theta}(\bm x^\mu)=\mathbb C,
\]
which should raise some serious concerns about the
physical interpretation of such coordinates under
a quantum perspective.
\subsection{Dangers of the Moyal expansion}
\label{subsec:misuses}
We see that we are facing a confusing situation. The same expression
\[
m\big(e^{(i/2)\theta^{\mu\nu}\partial_\mu\otimes\partial_\nu} f\otimes g\big)
\]
can be used for pairs $f$, $g$ of polynomials in \(\mathcal A_\theta\)
or of functions in
a dense sub algebra of \(\mathcal E_\theta\) (not containing the
polynomials); in both cases it gives the right
product in the corresponding algebra, which in both cases is denoted
by~\(\star\). However the two algebras are dramatically non isomorphic.

One common misuse is to understand \(\star\) as a product in
\(\mathcal A_\theta\) (for example, writing down \eqref{eq:not_innocent}),
yet making use of the Weyl quantisation prescription, which only makes sense
in \(\mathcal E_\theta\). The two are mutually
excluding, as we have seen. Formal computations under such a mixed formalism
are somewhat out of control, since one may inadvertently switch from
an algebra to the other, and back;
the result might depend on the way the computation
is done.

Let us see another example of what happens if one does not specify where the
symbols live:
\begin{quote}
{\itshape ``Let $f$, $g$ be functions with disjoint supports, and consider their Moyal
\(\star\)-product
\[
f\star g=
m\big(e^{(i/2)\theta^{\mu\nu}\partial_\mu\otimes\partial_\nu} f\otimes g\big).
\]
Since derivatives cannot enlarge the domains, at any order the product
vanishes, and \(f\star g=0\).''
}\end{quote}
In the best case, this statement is a trivial
tautology (whatever of the two algebras we choose to work with):
if two real-analytic symbols or two polynomials
have disjoint supports, at least one of the two is zero, so that their
product is evidently zero; which is certainly true, yet not so thrilling.

If $f$, $g$ are generic \(\mathcal C^\infty\)
functions, the above statement is meaningless,
since in general the Moyal expansion does not converge and there is no
\(f\star g\) at all. One solution is to understand the Moyal product
in the sense of formal power series in~\(\theta\); in this direction,
see e.g.~\cite{Aschieri:2009zz}, and references therein
(especially to work of Drinfel'd and Kontsevich).
However, doing this one misses something important from the point
of view of nonlocality.
Indeed, the integral form of the twisted product in \(\mathcal E_\theta\) is
intrinsically non local. On the contrary, the structure
of the Moyal expansion alone
is unable to reproduce the non local character of the full product: for generic
functions, it is non-commutative but local (look at the supports) at any order\footnote{By the way, this
also provides a proof that geometric quantisation is not equivalent to Weyl quantisation, since
the resulting algebras are deeply inequivalent.}.
Only when applied to real analytic {\itshape symbols} is the Moyal expansion
able to catch the full non local  content of the twisted
product; for, in the analytic world (as in the stock exchange),
local is global.

Finally, quite common a misuse is to formally use the Moyal expansion
as a ``product'' of functions on unspecif\/ied type, in the spirit
of ``existence in notation''. As it is easy to imagine, this also
is completely out of control. A spectacular example is the famous saga
of (non existing) ``violations of unitarity'',
which will be discussed in some detail in Section~\ref{sec:qft_qst}, and especially in Section~\ref{subsec:un_viol}.
We also will see that Filk diagrams and rules can be obtained in this way.

Let me conclude with a point of history. Weyl quantisation was suggested,
in the framework of canonical quantisation, by
Hermann Weyl in his famous book \cite{weyl:1928};
everybody interested in this
f\/ield should f\/ind the time for reading those beautiful pages.
Weyl also advocated the usefulness of symbolic calculus; however,
he did not write down the explicit formulas. The f\/irst appearance of the
deformed product is in the famous paper of von Neumann, on the uniqueness
of Schr\"odinger operators \cite{vonNeumann:1931}. However, for technical reasons,
he decided to work in Fourier space, so that he only wrote down the twisted
convolution product. Wigner found the solution of the inverse
problem (given the operator, f\/ind the corresponding symbol), while he was
working to his (unsuccessful)
theory of negative probabilities applied to quantum physics. His follower
Moyal wrote down (the quantum mechanical version of)
the tricky expansion we discussed in Section~\ref{subsec:moy_exp}. The correct integral formula for the twisted product
in conf\/iguration space f\/irst appeared in printed form  in papers by Baker and
Pool (see references of~\cite{Estrada:1989da}).
Finally, the covariant version of the Weyl quantisation (in terms of the
quantum coordinates of the spacetime, instead of the Schr\"odinger coordinates
of the phase space) and the corresponding Weyl calculus was f\/irst suggested
by Doplicher, Fredenhagen and Roberts~\cite{Doplicher:1994tu}.

\subsection{Existence of representations}
\label{subsec:exist_coord}
We shortly recall the very simple, yet ef\/fective argument of \cite{Doplicher:1994tu}
for the existence of the unbounded operators associated to the quantisation;
for more general choices of \(\theta\) and for time/space commutative models,
see Section~\ref{subsec:sad_fate}).
Let \(\varLambda\) be a \(4\times 4\) matrix (not necessarily in the Lorentz
group); let moreover\footnote{The matrix \(S\) was denote \(\sigma_0\) in
\cite{Doplicher:1994tu}. The change of notation is motivated by the fact that here we use
repeated sum over dummy indices instead of matrix notation.}
\[
S=({S}^{\mu\nu})=
\begin{pmatrix}
0&0&-1&0\\
0&0&0&-1\\
1&0&0&0\\
0&1&0&0
\end{pmatrix}.
\]
Then def\/ine
\begin{subequations}
\label{eq:basic_rep}
\begin{gather}
\bm X^0=\bm P_1,\\
\bm X^1=\bm P_2,\\
\bm X^2=\bm Q_1,\\
\bm X^3=\bm Q_2,
\end{gather}
\end{subequations}
where \(\bm P_j=-i\partial_j\) and \(\bm Q_j=s_j\cdot\) are the usual Schr\"odinger
operators on \(\mathfrak H=L^2(\mathbb R^2,ds_1ds_2)\)
for the particle on the plane,
fulf\/illing \([\bm P_j,\bm Q_k]=-i\delta_{jk}\bm I\) strongly
(this is a purely formal analogy, with no physical interpretation).
Then clearly the Weyl form of the relations
\[
[{\bm X}^\mu,{\bm X}^\nu]=i{S}^{\mu\nu}\bm I
\]
is fulf\/illed (as regular relations;
see last paragraph of Section~\ref{subsec:weyl_rels} at page
\pageref{subsec:weyl_rels}). It follows that, for any matrix
\(\varLambda=({\varLambda^\mu}_\nu)\) (not necessarily in the Lorentz group; but we use a covariant notation for later convenience),
the operators
\[
\bm x^\mu={\varLambda^\mu}_\nu\bm X^\nu
\]
fulf\/il
\[
[\bm x^\mu,\bm x^\nu]=i\theta^{\mu\nu}\bm I
\]
in the Weyl form~\eqref{eq:weyl_rels}, where
\[
\theta^{\mu\nu}=
{\varLambda^\mu}_{\mu'}{\varLambda^\nu}_{\nu'}{S}^{\mu'\nu'}.
\]
This gives
at once the class of matrices \(\theta\) for which the argument works,
and existence of the corresponding operators \(\bm x^\mu\).

Note also
that if in addition \(\varLambda\) is invertible then, given
the operators \(\bm x^\mu\), we may reconstruct the Schr\"odinger operators
$\bm P_j$, $\bm Q_j$. Hence in this case we also know that there cannot be other
irreducible representations (up to equivalence), as a corollary of
von Neumann uniqueness of Schr\"odinger operators~\cite{vonNeumann:1931}.

\subsubsection*{Translation covariant representations}
For the special case of \(\theta=S\), let us see how to construct a
translation covariant representation. The idea is to consider other two
independent Schr\"odinger pairs $\bm P_3$, $\bm Q_3$, $\bm P_4$, $\bm Q_4$,
so that \([\bm P_j,\bm Q_k]=i\delta_{jk}\bm I\), \(j,k=1,2,3,4\). Now take
\(\bm X^\mu\) as in \eqref{eq:basic_rep}; moreover, take
\begin{subequations}
\label{eq:basic_rep_momenta}
\begin{gather}
\bm \Pi_0=\bm Q_1+\bm Q_3,\\
\bm \Pi_1=\bm Q_2+\bm Q_4,\\
\bm \Pi_2=\bm P_3-\bm P_1,\\
\bm \Pi_3=\bm P_4-\bm P_2,
\end{gather}
\end{subequations}
and \(\bm\Pi^\mu=g^{\mu\nu}\bm\Pi_\nu\).
It is easy to check that
\[
[\bm \Pi^\mu,\bm \Pi^\nu]=0,\qquad[\bm \Pi^\mu,\bm X^\nu]=-ig^{\mu\nu}\bm I;
\]
hence the \(\bm \Pi^\mu\)'s are the pairwise commuting generators of
spacetime translations (in this parti\-cu\-lar representation). Note that the
localisation algebra has no dynamical content by itself,
so that these generators
can not be interpreted as the ``momenta'' of some theory.
A similar trick works for every invertible \(\theta\), see
\cite{Doplicher:1994tu}\footnote{There, the conjugate Hilbert space
is used to get compact formulas; note that if \((\bm J\psi)(s)=\overline{\psi(s)}\)
is the usual conjugation on \(L^2(\mathbb R,ds)\), then \(\bm{JPJ}=-\bm P\).}.

\section{Covariance and the DFR model}
\label{sec:dfr}
We now will describe the DFR model \cite{Doplicher:1994tu}. We will take a bottom-up
strategy. Firstly, we will try to cure the lack of covariance
by a simple construction, and obtain a class of $C^*$-algebras
which are covariant under the action of the full Poincar\`e group.
Then we will select a particular algebra in this class (corresponding
to a particular orbit of the antisymmetric tensors under the action of the
Lorentz group). Finally we will show that this algebra fulf\/ils physically
motivated uncertainty relations.

This line of exposition does not make justice to the original paper, where the
structure was {\itshape derived} from very general assumptions and ans\"atze
(top-down approach), instead of being constructed by  hand. However I
hope that  this will make easier, for a broad readership,
to grasp in a few pages some of the
more delicate technical aspects. Moreover, the comparison with the usual
approach to the so called ``canonical quantum spacetime'' will be more
transparent.  The price to pay is that some apparently
arbitrary assumptions will have to f\/ind their motivations only in the end.

In what follows, we will assume that \(\theta\) is such that there
exists a unique representation of the Weyl relations
\eqref{eq:weyl_rels}. Moreover, we will
keep separate the purely algebraic structure
and the choice of the length scale (or unit of measure) by setting
\[
\theta^{\mu\nu}=\lambda^2{\sigma}^{\mu\nu},
\]
where \(\lambda\) is a length and the entries of \(\sigma\) are
pure numbers.

Correspondingly, we will write
\(\star_{\sigma}\)
instead of \(\star\) for the product in \(\mathcal E_\sigma\);
the symbol \(\star\) will be reserved
for a more general product. Hence,
we may rewrite \eqref{eq:tw_prod_expl} as
\begin{equation}
(f\star_{\sigma}g)(x)=\int dk\; e^{ik_\mu x^\mu}
\int dh\;\check f(h)\check g(k-h)
e^{-i\frac{\lambda^2}{2}
h_\mu{\sigma}^{\mu\nu}k_\nu}.
\end{equation}

\subsection{``Be wise, covariantise''}
\label{subsec:wise}
For a given \(L=(\varLambda,a)\) in the Poincar\'e group \(\mathscr P\),
let
\begin{gather*}
f'(x)=f(\varLambda^{-1}(x-a)),\quad\quad g'(x)=g(\varLambda^{-1}(x-a)),
\\
(f\star_{\sigma} g)'(x)=(f\star_{\sigma} g)(\varLambda^{-1}(x-a)).
\end{gather*}
Of course, the product \(\star_\sigma\) is not covariant: in general, we have
\[
f'\star_\sigma g'\neq (f\star_\sigma g)'.
\]
However (as one can easily imagine)
a simple computation yields
\[
(f'\star_{\sigma'}g')(x)=(f\star_{\sigma} g)'(x),
\]
where
\[
{\sigma'}^{\mu\nu}={\varLambda^\mu}_{\mu'}{\varLambda^\nu}_{\nu'}
{\sigma}^{\mu'\nu'}.
\]
This suggests a very simple way out. The idea is to turn the parameter
\(\sigma\) into a variable. Let~\(\Sigma\) be a set of antisymmetric
matrices which is stable under the action
\[
\varLambda:\sigma\mapsto \sigma'=({\varLambda^\mu}_{\mu'}{\varLambda^\nu}_{\nu'}
{\sigma}^{\mu'\nu'}).
\]
Then consider generalised symbols, namely
functions of \(\Sigma\times\mathbb R^4\), and def\/ine
the product
\[
(f\star g)(\sigma;x)=
\int dk\; e^{ik_\mu x^\mu}
\int dh\;\check f(\sigma;h)\check g(\sigma;k-h)
e^{-i\frac{\lambda^2}{2}
h_\mu{\sigma}^{\mu\nu}k_\nu},
\]
where of course \(\check f\)  is the Fourier
transform of \(f(\sigma;\cdot)\) for each \(\sigma\) f\/ixed.
In addition, set pointwise complex conjugation
as the involution:
\[
f^\star(\sigma;x)=\overline{f(\sigma;x)}.
\]
The above gives a well def\/ined $*$-algebra, which we denote \(\mathcal E^{(0)}\).

If we def\/ine the action of \(L=(\varLambda,a)\in\mathscr P\),
the Poincar\'e group, by
\begin{gather*}
(\gamma(L)f)(\sigma';x')=\det\varLambda f(\sigma;x),\\
{\sigma'}^{\mu\nu}={\varLambda^\mu}_{\mu'}{\varLambda^\nu}_{\nu'}
\sigma^{\mu'\nu'},\qquad x'=\varLambda x+a,
\end{gather*}
then by construction we obtain covariance:
\[
\gamma(L)(f\star g)=
(\gamma(L)f)\star (\gamma(L)g).
\]
In more technical language, for each \(L\), \(\gamma(L)\) is a $*$-automorphism
of \(\mathcal E^{(0)}\), and \(\gamma(L_1)\gamma(L_2)=
\gamma(L_1L_2)\); this situation is often described in the mathematical
literature as \(\gamma\) providing an action of \(\mathcal \mathscr P\)
by automorphisms of \(\mathcal E^{(0)}\).

We still have to f\/ix the class of functions, and the set \(\Sigma\).
Of course, for every \(\sigma\) f\/ixed we have to require that
\(f(\sigma;\cdot)\) is an admissible symbol.
As for \(\Sigma\), the simplest choice is to pick a single orbit
in the space of antisymmetric matrices under the given action.
For the moment this choice only can be motivated by the quest for
simplicity;
but we will see that it corresponds to assume that there is one only
characteristic length driving the algebraic structure.

For the purpose of computations, the above is almost all one needs. However,
since we put emphasis on the importance of existence of $C^*$-completions
(especially in order to have a~non trivial theory from the spectral point
of view), let us give a hint about the underlying $C^*$ structure,
and how to unveil it.

Giving a closer look at the way the product of generalised symbols
is def\/ined, we may see that there is a natural bundle structure emerging:
indeed, the product
\[
(f\star g)(\sigma;\cdot)=f(\sigma;\cdot)\star_\sigma g(\sigma;\cdot),
\qquad\sigma\in\Sigma,
\]
appears as a f\/ibrewise product over the base space \(\Sigma\).
This suggest to take \(f\) as a continuous
function (vanishing at inf\/inity) sending each \(\sigma\in\Sigma\)
to an element of \(\mathcal E_\sigma\), namely a continuous section of a bundle
of algebras. At a f\/irst glance,
it seems to be a non trivial bundle, since each f\/ibre
is dif\/ferent; however we now can remember that every f\/ibre is dense in the same
$C^*$-completion, which is the algebra \(\mathcal K\) of compact operators
on the separable Hilbert space. Hence
we are naturally led to consider a trivial bundle of $C^*$-algebras, with
base space \(\Sigma\) and standard f\/ibre \(\mathcal K\). It may be shown
(see \cite{Doplicher:1994tu}) that actually
\[
\mathcal E=\mathcal C_0(\Sigma,\mathcal K)
\]
is the unique $C^*$-completion
of the algebra \(\mathcal E^{(0)}\)
of continuous sections ($=$~generalised symbols).

Note that the action \(\gamma\) of the Poincar\'e group on the algebra of
continuous
sections maps each f\/ibre over \(\sigma\) onto the f\/ibre over \(\sigma'\)
and extends by continuity to an action by automorphisms
of the full $C^*$-algebra \(\mathcal E\).

\subsection{Representations}
\label{subsec:dfr_full_rep}
In this section, we describe how to construct the representations
of the $C^*$-algebra \(\mathcal E=\mathcal C_0(\Sigma,\mathcal K)\),
using the Dirac notation; for a mathematically rigorous treatment,
see the original paper.

Let us f\/ix \(\Sigma\) to be the orbit containing the matrix
\(S\) described at the end of Section~\ref{sec:which_algebra}.
This is not technically
necessary at this point of the discussion, but it simplif\/ies the
notations, and we will anyhow f\/ind that this choice is the correct one.
We also f\/ix the choice \(\lambda=\lambda_P\), the Planck length.

The reader should have clear in mind the idea underlying the
construction described in Section~\ref{subsec:exist_coord}.

We consider the Hilbert space\footnote{If there were a Lorentz invariant
measure \(d\sigma\) on \(\Sigma\), we could have taken kets of the form
\(\ket\sigma\ket\xi\) and integrate with \(d\sigma\) instead
of \(d\varLambda\). It is a fact of life that such a measure does not exist.}
\[
\mathfrak H=L^2(\mathscr L,d\varLambda )\otimes L^2(\mathbb R^2,ds_1ds_2),\]
where \(d\varLambda\) is the Haar measure of the Lorentz group \(\varLambda\).
As usual, we associate to it a complete set of generalised kets
\[
\ket{\varLambda}\ket{s_1,s_2},\qquad
\varLambda\in\mathscr L,(s_1,s_2)\in\mathbb R^2,
\]
with bracket
\[
\{\bra{\varLambda}\bra{s_1,s_2}\}\{\ket{\varLambda'}\ket{s'_1,s'_2}\}=
\bracket{\varLambda}{\varLambda'}\bracket{s_1,s_2}{s_1',s_2'}=
\delta(\varLambda^{-1}\varLambda')\delta(s_1-s_1')\delta(s_2-s_2'),
\]
where integrals are taken with the measure \(d\varLambda ds_1ds_2\) and
\(\delta(\varLambda)d\varLambda\) is the purely atomic normalised
measure on \(\mathscr L\), concentrated on \(I\).

We def\/ine the operators \(\bm q^\mu\) by their actions on the kets
\(\ket{\varLambda}\ket{\xi}\):
\[
\bm q^\mu\ket{\varLambda}\ket{\xi}=\lambda_P\ket{\varLambda}
\{{\varLambda^\mu}_\nu\bm X^\nu\ket\xi\},
\]
where the \(\bm X^\mu\) are described in Section~\ref{subsec:exist_coord}. These operators being def\/ined in terms of the
Schr\"odinger operators, we have
\[
\text{spec}(\bm q^\mu)=\mathbb R.
\]

We may easily check that the operators \(\bm Q^{\mu\nu}\) def\/ined by
\[
[\bm q^\mu,\bm q^\nu]=i\lambda_P^2\bm Q^{\mu\nu}
\]
are simultaneously diagonalised by the kets
\(\ket{\varLambda}\ket{s_1,s_2}\):
\[
\bm Q^{\mu\nu}\ket{\varLambda}\ket{\xi}=
{\varLambda^\mu}_{\mu'}{\varLambda^\nu}_{\nu'}{S}^{\mu'\nu'}
\ket{\varLambda}\ket{\xi}.
\]
It follows that the spectrum of each \(\bm Q^{\mu\nu}\) is
\[
\text{spec}(\bm Q^{\mu\nu})=\{\sigma^{\mu\nu}:\sigma\in\Sigma\}=\mathbb R.
\]

We have a unitary representation \(\bm U\) of the Lorentz group
\[
\bm U(\varLambda)\ket M\ket{s_1,s_2}=\ket{\varLambda M}\ket{s_1,s_2};
\]
it fulf\/ils
\begin{gather*}
\bm U(\varLambda)^{-1}\bm q^\mu \bm U(\varLambda)={\varLambda^\mu}_\nu\bm q^\nu,\\
\bm U(\varLambda)^{-1}\bm Q^{\mu\nu}\bm U(\varLambda)=
{\varLambda^\mu}_{\mu'}{\varLambda^\mu}_{\mu'}\bm Q^{\mu'\nu'}.
\end{gather*}
Let now \(f=f(\sigma;x)\) be a generalised symbol, as described in Section~\ref{subsec:wise}. We may def\/ine its DFR quantisation {\itshape \`a la Weyl}
\[
f(\bm Q;\bm q)\ket{\varLambda}\ket{\xi}=
\ket{\varLambda}\left\{\int dk\;\check f(({\varLambda^\mu}_{\mu'}
{\varLambda^\nu}_{\nu'}S^{\mu'\nu'});k)
e^{i\lambda_Pk_\mu{\varLambda^\mu}_\nu\bm X^\nu}\ket{\xi}\right\}.
\]
Note that if we take a function of the form
\[
f(\sigma;x)=(f_1\otimes f_2)(\sigma;x)=f_1(\sigma)f_2(x),\]
then the DFR quantisation prescription gives the operator product
\[
(f_1\otimes f_2)(\bm Q;\bm q)=f_1(\bm Q)\int dk\check f_2(k)e^{ik\bm q},
\]
where \(f_1(\bm Q)\) is the usual evaluation of a function on a set of pairwise
commuting operators (joint functional calculus).
By a direct check, we f\/ind in general
\[
f(\bm Q;\bm q)g(\bm Q;\bm q)=(f\star g)(\bm Q;\bm q),
\qquad \bar f(\bm Q;\bm q)=f(\bm Q;\bm q)^*
\]
and
\[
\bm U(L)f(\bm Q;\bm q)\bm U(L)^{-1}=(\gamma(L)f)(\bm Q;\bm q),\qquad
L\in\mathscr P.
\]

Finally, we may consider the functions \(\sigma^{\mu\nu}\sigma_{\mu\nu}\)
and \((\sigma^{\mu\nu}(\ast\sigma)_{\mu\nu})^2\) of \(\Sigma\), where
\((\ast \sigma)_{\mu\nu}=\tfrac 12\epsilon_{\mu\nu\rho\tau}\sigma^{\rho\tau}\) is the Hodge dual of \(\sigma\). These functions
are invariants of the full Lorentz group, and there are no other invariants
independent from these two. Since they are invariants, their values are constant
and can be computed at any point in \(\Sigma\), e.g.\ at \(S\). Hence{\samepage
\[
\sigma^{\mu\nu}\sigma_{\mu\nu}=0,\qquad
(\sigma^{\mu\nu}(\ast\sigma)_{\mu\nu})^2=16,\qquad\sigma\in\Sigma.
\]
It follows that the same relations hold true with \(\bm Q\) in the
place of \(\sigma\).}

The Lorentz covariant representation \((\bm q^\mu,\bm U)\) described above
is essentially (i.e.\ up to equiva\-lence and multiplicity) unique.
This follows from a ``f\/ibrewise'' von Neumann uniqueness
argument\footnote{The reader may recognise in the above construction a direct
integral over \(\mathscr L\) of irreducible representations labeled by
\(\sigma=\varLambda S\varLambda^t\). For generic kets
\(\ket\psi\in\mathfrak H\) of f\/inite length,
the functions \(\psi(\varLambda;s_1,s_2)=
\{\bra\varLambda\bra{s_1,s_2}\}\ket\psi\) may be recognised as the measurable
f\/ields \(\varLambda\mapsto\psi(\varLambda;\cdot)\) in the direct
integral \(\mathfrak H=\int^\oplus d\varLambda L^2(\mathbb R^2)\).}
\cite{Doplicher:1994tu}.

An adaptation of the same argument, where now we take the translation covariant
representations described at the end of Section~\ref{subsec:exist_coord}
as building blocks, will give a full Poincar\'e covariant representation.

\subsection{The DFR model recovered}
\label{sec:dfr_recovered}
In the preceding section we have shown that the universal, covariant
representation of the
$C^*$-algebra \(\mathcal C_0(\Sigma,\mathcal L)\) is associated with
selfadjoint operators \(\bm q^\mu\) fulf\/illing
\begin{subequations}
\label{eq:dfr}
\begin{equation}\label{eq:dfr_1}
[\bm q^\mu,\bm q^\nu]=i\lambda_P^2\bm Q^{\mu\nu},
\end{equation}
where
\begin{gather}
\label{eq:dfr_2}
[\bm q^\mu,\bm Q^{\nu\rho}]=0,\\
\label{eq:dfr_3}
\bm Q^{\mu\nu}\bm Q_{\mu\nu}=0,\\
\label{eq:dfr_4}
\left(\bm Q^{\mu\nu}(\ast\bm Q)_{\mu\nu}\right)^2=16\bm I.
\end{gather}
\end{subequations}
In \cite{Doplicher:1994tu},
the above relations were the starting point for the analysis.
Indeed, the pattern proposed there was precisely opposite to the one we
followed.

So, everything is a consequence of the relations~\eqref{eq:dfr}
and their full Poincar\'e covariance expressed by
\begin{subequations}
\label{eq:dfr_cov}
\begin{gather}
\bm U(\varLambda,a)^{-1}\bm q^\mu \bm U(\varLambda,a)={\varLambda^\mu}_\nu\bm q^\nu+a^\mu \bm I,\\
\bm U(\varLambda,a)^{-1}\bm Q^{\mu\nu}\bm U(\varLambda,a)=
{\varLambda^\mu}_{\mu'}{\varLambda^\mu}_{\mu'}\bm Q^{\mu'\nu'},
\end{gather}
\end{subequations}
for \((\varLambda,a)\in\mathscr P\), the Poincar\'e group.

But where
do they come from? To give a quantitative content to the remarks of
Section~\ref{sec:motivations}, in \cite{Doplicher:1994tu} some important steps were made
precise.
\begin{itemize}\itemsep=0pt
\item The (limited)
scope of the model: an idealised situation describing scattering
experiments in particle physics. This means that the density of processes is
so low that it does not produce signif\/icant deviations from the f\/lat geometry
of the laboratory; however their energy is so high that deviations
from classical f\/lat spacetime are produced only at very small scales.
\item The necessity of consistence with usual particle physics (described
in terms of quantum f\/ields). This requires in particular that the usual
classical (commutative) Minkowski spacetime must be obtained in the large
scale limit (possibly at the cost of extra dimensions).
\item The necessity of ordinary Poincar\'e covariance (phrased in the dear old
Wigner's language for quantum symmetries). This for many reasons:
a) the need for preserving Wigner's classif\/ication of
particles scattered far away (which are then detected in the
``classical region'', where non-commutativity is smeared away:
in- and out-f\/ields are def\/ined on the classical spacetime); b) the remark that
transformations of reference frames are global and af\/fect the cosmic scale as
well as the Planck scale; c) the need for preserving
the symmetry between active and passive point of view (think of an experiment
in a laboratory).
\end{itemize}
Within this conceptual framework, we may formulate the f\/irst DFR problem:
\begin{quote}(DFR1)
{\itshape Find relations between the uncertainties \(\Delta x^\mu\), which
prevent the formation of trapped surfaces enclosing the observed region,
as an effect of localisation alone.}
\end{quote}
Note that, in view of the limited scope of the model,
dynamical black hole formation is prevented from
the outset, since the large scale geometric background is f\/ixed, and it is
f\/lat. In a would-be, still unknown more general model, one should be able
to distinguish dynamical black hole formation from localisation-induced
black hole formation, and f\/ind conditions preventing the latter only.

It is possible to carry over a heuristic analysis,
where localisation states are modeled by coherent
states of the form \(e^{i\phi(f)}\ket 0\) (\(\ket 0\) is the Fock vacuum
for a free f\/ield \(\phi\)). If the (``commutative'') test function \(f\) has
support in a 4-box of sides \(\Delta x^\mu\), bounds
from Einstein equations are obtained
by (crude) estimates on the expectation of the energy-momentum tensor
in such states. We omit the details, for which the reader is referred to
the original paper \cite{Doplicher:1994tu}. This leads to the relations
\begin{subequations}
\label{eq:stur}
\begin{gather}
\Delta x^0(\Delta x^1+\Delta x^2+\Delta x^3)\geqslant \lambda_P^2,\\
\Delta x^1\Delta x^2+\Delta x^1\Delta x^3+
\Delta x^2\Delta x^3
\geqslant \lambda_P^2.
\end{gather}
\end{subequations}
The second DFR problem comes from the idea of following the pattern which led
to canonical quantisation in the 1920s: given the uncertainty relations,
f\/ind the commutation relations which reproduce them by the uncertainty
theorem\footnote{We recall that, given two selfadjoint operators
$\bm A$, $\bm B$, then
for any state \(\omega\) (which may be a vector state or a~density matrix, it is
irrelevant), we have \(\Delta_\omega(\bm A)\Delta_\omega(\bm B)\geqslant
\frac 12|\omega([\bm A,\bm B])|\).}.
\begin{quote} (DFR2) {\itshape Find commutation relations between
selfadjoint operators
\(\bm q^\mu\) such that the relations \eqref{eq:stur}
are fulfilled if one replaces
\(\Delta_\omega \bm q^\mu\) for \(\Delta x^\mu\), no matter which
state \(\omega\) is chosen.}
\end{quote}
Above,
\(\Delta_\omega \bm q^\mu\) is the usual uncertainty of the operator
\(\bm q^\mu\) in the state \(\omega\), namely
\[
\Delta_\omega\bm q^\mu=\sqrt{\omega((\bm q^\mu)^2)-\omega(\bm q^\mu)^2}.
\]

Now it is clear that there is nothing bad in the
uncertainty relations not having a covariant look'n feel since,
despite the misleading notation, in general
the uncertainties \(\Delta_\omega \bm q^\mu\)  are {\itshape not}
the components of a covariant vector. Indeed \(\Delta_\omega\) is
{\itshape not} a linear functional on operators, so that
\({\varLambda^\mu}_\nu\Delta_{\omega}(\bm q^\nu)\neq
\Delta_{\omega}({\varLambda^\mu}_\nu\bm q^\nu)\), in general. What is important
is to check that the relations hold in any reference frame.

To solve problem DFR2, we observe that a measure of non-commutativity
of \(n\) elements is given by the determinant
\[
[\bm A_1,\bm A_2,\dotsc,\bm A_n]:=\sum\epsilon_{j_1\dots j_n}\bm A_{j_1}\dotsm \bm A_{j_n}=
\det
\begin{bmatrix}
\bm A_1&\bm A_2&\dotsc&\bm A_n\\
\bm A_1&\bm A_2&\dotsc&\bm A_n\\
\vdots&\vdots&\ddots&\vdots\\
\bm A_1&\bm A_2&\dotsc&\bm A_n
\end{bmatrix}
\]
which for \(n=2\) reduces to the usual commutator. Hence a model with
\(4\) coordinates is f\/ixed by giving conditions on
\begin{equation}\label{eq:determinants}
[\bm q^\mu,\bm q^\nu],\qquad [\bm q^\mu,\bm q^\nu,\bm q^\rho],
\qquad [\bm q^\mu,\bm q^\nu,\bm q^\rho,\bm q^\tau].
\end{equation}
Setting
\[
(\ast \bm Q)_{\mu\nu}=\frac 12\epsilon_{\mu\nu\rho\tau}\bm Q^{\rho\tau},
\]
we have
\[
[\bm q^0,\dotsc,\bm q^3]=-\frac 12 \bm Q^{\mu\nu}(\ast\bm Q)_{\mu\nu}
\]
which is invariant under the special Lorentz group \(\mathscr L^\uparrow_+\).
There also is the full Lorentz invariant
\[
\bm Q^{\mu\nu}\bm Q_{\mu\nu}
\]
and there are no other invariants which are independent from these ones.
Hence the only Poincar\'e invariant constrains which can be formulated with
the determinants \eqref{eq:determinants} must be given in terms of these
objects and of \([\bm q^\mu,\bm q^\nu,\bm q^\rho]\).

With the simplifying ansatz \([\bm q^\mu,\bm Q^{\nu\rho}]=0\)
(no physical motivations for it), which may be equivalently written as
\[
[\bm q^\mu,\bm q^\nu,\bm q^\rho]=0,
\]
it is possible to show that the solution of problem DFR2 is provided precisely
by \eqref{eq:dfr}.

Writing down commutation relations is not enough! Existence of operators
actually fulf\/illing them is not granted; hence we have the third DFR problem
\begin{quote}
(DFR3) {\itshape Find operators \(\bm q^\mu\) which fulf\/il the relations
\eqref{eq:dfr}, and are covariant.}
\end{quote}
This problem is solved precisely by the construction described
in Section~\ref{subsec:dfr_full_rep}.

We close by observing that, in the DFR model, the spectrum of
\([\bm q^0,\bm q^1,\bm q^2,\bm q^3]\) is \(\{\pm 2\lambda_P^4\}\),
while the spectrum of \([\bm q^\mu,\bm q^\nu][\bm q_\mu,\bm q_\nu]\)
is zero. Hence in this model there is one only universal length,
which is precisely \(\lambda_P\).

\subsection{Working with the DFR model}
\label{subsec:working}
Due to the uniqueness of the representation, it is useful to forget the
dif\/ference between abstract elements and their representations. So,
we think of \(f(\bm Q;\bm q)\) both as of an operator, and of
an element of the $C^*$-algebra; and, since the algebra of
continuous sections is naturally embedded in its unique $C^*$-completion,
we are allowed to think of \(f(\bm Q;\bm q)\) as a function of \((\sigma,x)\).
This way of reasoning may seem at f\/irst
rather confusing, since one thinks of the same symbol as of an operator,
a function, or an element of an abstract $C^*$-algebra depending on the
convenience of the moment. However, it is quite a fruitful point of view.

To make this clear, let us work out some example which will be used in
later  sections.

Firstly, observe that, if \(f=f(x)\) does not depend on \(\sigma\),
\[
f(\bm q)=\int dk \check f(k)e^{ik_\mu\bm q^\mu}
\]
is well def\/ined, yet it does not belong to the spacetime algebra
\(\mathcal E=\mathcal C_0(\Sigma,\mathcal K)\) since \(f\)
does not vanish at inf\/inity as a function of \(\sigma\). However,
\(f(\bm q)\) is a multiplier of the algebra,
in the sense that \(f(\bm q)g(\bm Q;\bm q)\) is in \(\mathcal E\)
for every admissible symbol \(g\). This can be understood also without knowing
the abstract def\/inition of multipliers algebra of a $C^*$-algebra\footnote{This generalises a very simple situation: a bounded continuous function
times a continuous function vanishing at inf\/inity vanishes itself at inf\/inity,
hence \(\mathcal C_b(\mathbb R^4)\) is the multipliers algebra of
\(\mathcal C_0(\mathbb R^4)\). It is the biggest algebra of functions with
this property and thus it is a canonical object.}.

{\sloppy The functions of \(\sigma\) alone also are natural multipliers: if
\(f=f(\sigma)\) is continuous and bounded and \(g=g(\sigma;x)\) is an admissible
symbol, then \(f(\sigma)g(\sigma;x)\) is an admissible symbol and
\(f(\bm Q)g(\bm Q;\bm q)\) has an unambiguous meaning. Moreover, \(f(\bm Q)\)
is a central multiplier:
\(f(\bm Q)g(\bm Q;\bm q)=g(\bm Q;\bm q)f(\bm Q)\); or, if we enjoy
using exact terms, we may say that
\(f(\bm Q)\) belongs to the centre \(Z(M(\mathcal E))\)
of the multipliers algebra \(M(\mathcal E)\) of \(\mathcal E\). Note also that
we have the natural identif\/ication,
\[
Z(M(\mathcal E))=\mathcal C_b(\Sigma),
\]
the algebra of bounded continuous functions
of \(\Sigma\).

}

The algebra \(\mathcal E\) has no unit, because the only candidate would
be, seen as a section, the constant function \(\sigma\mapsto 1\); which does
not vanish at inf\/inity as a~function of \(\sigma\),
and is not \(L^1\) as a~function of \(x\) for
any f\/ixed \(\sigma\). If we ask the same question while looking at the
abstract $C^*$-algebra  \(\mathcal E=\mathcal C_0(\Sigma,\mathcal K)\),
we see that the only candidate to be the identity would be the constant function
\(\sigma\mapsto \bm 1\); which does not vanish at inf\/inity as a function of
\(\sigma\),  and \(\bm 1\) is not compact. Of course, both candidates coincide
with the unit of the multipliers algebra, up to natural identif\/ication.

Let us now consider the map
\[
f(\bm Q;\bm q)\mapsto \int dx\;f(\bm Q;x)\delta(x^0-t).
\]
For every f\/ixed \(t\) this object depends on \(\bm Q\) only, and is thus
a central multiplier. The above
map extends to a map from the whole $C^*$-algebra \(\mathcal E\) to
the central multipliers.

This map has a very interesting feature: it is positive, namely it sends
positive elements into positive elements:
\[
\int dx\;f(\bm Q;x)^*f(\bm Q;x)\delta(x^0-t)\geqslant 0.
\]
Since \(f(\bm Q;x)^*f(\bm Q;x)=(\bar f\star f)(\bm Q;x)\), the above
can be written as
\[
\iint dx\;(\bar f\star f)(\bm Q;x)\delta(x^0-t)\geqslant 0.
\]
But central multipliers are identif\/ied with functions of \(\sigma\),
where positivity is just pointwise (in~\(\sigma\)) positivity. Hence
the above is equivalent to
\[
\int dx\;(\bar f\star f)(\sigma;x)\delta(x^0-t)\geqslant 0,
\qquad \sigma\in\Sigma.
\]
In technical terms, the above map is a conditional expectation, since in
addition it maps
\[
f(\bm Q)g(\bm Q;\bm q)\mapsto
f(\bm Q)\int dx\;g(\bm Q;x)\delta(x^0-t).
\]
In \cite{Doplicher:1994tu}, the following suggestive notation was proposed for
the above map:
\begin{equation}\label{eq:pos_exp}
\int_{q^0=t} d^3q\;f(\bm Q;\bm q):=\int dx\;f(\bm Q;x)\delta(x^0-t).
\end{equation}
Note that positivity {\itshape with respect to the $*$-algebra}
is important, it essentially means compatibility
with the uncertainty relations. For example, for every f\/ixed
\(a\in\mathbb R^4\), the map \(f(\bm Q;\bm q)\mapsto f(\bm Q;a)\) (evaluation at a point)
is not positive, since there's no reason why \((\bar f\star f)(\sigma,a)\)
should def\/ine a positive function of \(\sigma\); on the contrary, it cannot be positive, for otherwise it would allow to
construct (by further integrating over \(\sigma\))
an admissible localisation state which
would violate the uncertainty relations.
The physical interpretation of positivity of the conditional expectation~\eqref{eq:pos_exp} is that localisation at sharp time can be compensated
by complete space delocalisation.

The other important notion is that of derivative. Since we have the
action of translations, we may {\itshape define}
\(\partial f(\bm q)/\partial \bm q^\mu\) by
\[
f(\bm q+\varepsilon)= f(\bm q)+\varepsilon^\mu\frac{\partial f(\bm q)}{d\bm q^\mu}+
o(\varepsilon).
\]
A short computation shows that it is precisely
the same as the quantisation of \(\partial_\mu f\):
\[
\frac{\partial f(\bm q)}{d\bm q^\mu}=(\partial_\mu f)(\bm q).
\]

Finally, a comment on the states. Let \(\omega\) be a state on \(\mathcal E\),
namely a liner functional on \(\mathcal E\) which is positive and normalised.
Examples can be obtained by thinking of \(f(\bm Q;\bm q)\) as of an operator:
vector states\footnote{Since the def\/ining representation is reducible,
`vector state' is not the same as `pure state'.}
\[
f(\bm Q;\bm q)\mapsto\frac{\bra\xi f(\bm Q;\bm q)\ket\xi}{\bracket{\xi}{\xi}}
\]
and more general states of the form
\[
f(\bm Q;\bm q)\mapsto\text{Tr}(\rho f(\bm Q;\bm q)),
\]
where \(\rho\) is a positive operator with trace 1. A (regular) state can be
extended to the multipliers. Since the central multipliers are
an algebra of functions, then for any
(regular) state \(\omega\) there is a measure \(\mu_\omega\) on \(\Sigma\)
such that, for any central multiplier \(f(\bm Q)\),
\[
\omega(f(\bm Q))=\int d\mu_\omega(\sigma)f(\sigma).
\]

A state is said with optimal localisation if it minimises
\[
\sum_\mu\Delta_\omega(\bm q^\mu)^2.
\]
This quantity is not Lorentz invariant, hence this def\/inition depends on the
observer, and the pull back of such a state to a dif\/ferent Lorentz frame
does not have optimal localisation, in general.
Such states exist and are essentially given in terms of the quantum
mechanical coherent states. It can be shown that, if \(\omega\) has optimal
localisation, then the associated measure \(\mu_\omega\) is supported
by the orbit \(\Sigma_1\subset\Sigma\)
of the standard simplectic matrix \(S\) under the
space-orthogonal subgroup of \(\mathscr L\). See \cite{Doplicher:1994tu} for more details.

\subsection{Classical limit}

It is clear that, when \(\lambda_P\) goes to zero, then
\(f\star g\) goes to the pointwise product in the \(x\) variable, f\/ibrewise in
the variable \(\sigma\):
\[
(f\star g)(\sigma;x)\underset{\lambda_P\rightarrow 0}{\longrightarrow}
f(\sigma;x)g(\sigma;x).
\]
The classical limit of the DFR model, then, is the same as the large dilations
limit, and is
\(\mathbb R^4\times\Sigma\), the usual Minkowski
spacetime, times a manifold carrying
extra dimensions. This comment may be made rigorous using Rief\/fel's theory
of deformations \cite{rieffel-lincei}.

The manifold \(\Sigma\) has two connected components\(\Sigma_\pm\),
each of which
is homeomorphic to the tangent space of the 2-sphere, so that there are
four extra dimensions\footnote{This can be seen directly, by observing that
an antisymmetric 2-tensor has six independent entries, and there are two independent invariants def\/ining the manifold \(\Sigma\).}.
Hence the classical  limit can be described as
\(\mathbb R^4\times TS^2\times\{\pm 1\}\).

An unpleasant fact is that the extra dimensions are not compact, and thus
in principle could be observed at the macroscopic scale. One way out would be
to restrict the representation to the states which, in the large scale limit,
will become sharp points; namely the states with optimal localisation.
We recall that states with optimal localisation, restricted to the centre,
are supported by \(\Sigma_1\), which, under the homeomorphism
\(\Sigma_\pm\rightarrow TS^2\), is mapped precisely to the 2-sphere.
Hence in this case, the classical limit would be
\(\mathbb R^4\times S^2\times\{\pm 1\}\), with compact extra dimensions.
Unfortunately, the price to pay is the breakdown of Lorentz covariance, so that
dif\/ferent observers, connected by a Lorentz boost, would disagree on the
classical limit.

\subsection{Towards quantum geometry}
\label{subsec:quantum_geom}
At its most basic level,
classical geometry describes the relations of families of points.

A f\/irst step towards a generalisation of classical geometry is to discuss
independent events. Two quantum quantities are statistically independent if they
commute; a natural way for constructing independent copies of the
same observable \(\bm A\) is by means of tensors product:
$\bm A_1=\bm A\otimes \bm I$, $\bm A_2=\bm I\otimes \bm A$. We may
follow this pattern for def\/ining the quantum coordinates of~\(n\)
independent events
\[
\bm q_j^\mu=\underbrace{\bm I\otimes\dotsm
\overset{
\overset{\text{\footnotesize\(j^{\text{th}}\) slot}}{\downarrow}}%
{\bm I\otimes\bm q^\mu\otimes\bm I}\otimes \dotsm\otimes \bm I}_{\text{\footnotesize \(n\) factors}},\qquad j=1,\dotsc,n.
\]
Then the quantisation of a function \(f=f(x_1,\dotsc,x_n)\) of many classical
independent events is given by
\[
f(\bm q_1,\dotsc,\bm q_n)=
\int_{\mathbb R^{4n}} dk_1\dotsm dk_n\;\check f(k_1,\dotsc,k_n)
e^{i(k_1\bm q_1+\dotsb k_n\bm q_n)},
\]

However, there are two natural notions of tensor product on the spacetime
$C^*$-algebra. The usual one, which gives \(\mathcal E\otimes\mathcal E\simeq
\mathcal C_0(\Sigma\times\Sigma,\mathcal K)\), namely functions of two independent \(\sigma\)'s, and the f\/ibrewise tensor product, according to
which the tensor product of two sections is taken f\/ibrewise. The abstract
algebraic method for doing this is to understand \(\otimes\) as
a tensor product of \(Z\)-modules, where \(Z\) is the centre of the
multipliers algebra (see Section~\ref{subsec:working}). Essentially,
this amounts
to identify the commutators of independent events:
\[
[\bm q_j^\mu,\bm q_k^\nu]=i\lambda_P^2\delta_{jk}\bm{\mathcal Q}^{\mu\nu},
\]
where
\begin{equation}\label{eq:Q_the_same}
\bm{\mathcal Q}^{\mu\nu}= \bm Q^{\mu\nu}\otimes\bm I\otimes\dotsm\otimes\bm I=
\bm I \otimes\bm Q^{\mu\nu}\otimes\dotsm\otimes\bm I=\dotsb=
\bm I \otimes\bm I\otimes\dotsm\otimes\bm Q^{\mu\nu}.
\end{equation}

That this is the natural notion of tensor product  is conf\/irmed by the remark
that the separations between two independent events has the same commutation
relations of the basic coordinates, up to a factor: if we set
\[
\delta_{jk}\bm q=\bm q_j-\bm q_k,
\]
we f\/ind
\[
[\delta_{jk}\bm q^\mu,\delta_{jk}\bm q^\nu]=2\lambda_P^2\bm{\mathcal Q}^{\mu\nu},
\qquad j\neq k,
\]
where the right hand side does not depend on $j$, $k$ \cite{piac_tesi,Bahns:2003vb}.
Now the square Euclidean
distance \(\sum_\mu(\delta_{jk}\bm q^\mu)^2\) is bounded below
by\footnote{This can be seen directly in any state which is determined
on \(\bm{\mathcal Q}\) with expectation \(\sigma\in\Sigma_1\); in this case,
we only see the irreducible component corresponding to the special
representation described in Section~\ref{subsec:exist_coord}, in which case the
square Euclidean distance of two independent events is twice the Hamiltonian
of a four dimensional Harmonic oscillator. See \cite{Doplicher:1994tu}
for a proof that this is optimal.} \(4\lambda_P^2\); hence there is a natural
minimal Euclidean distance between independent events \cite{4-volume}.

Another pleasant feature which follows from adopting the f\/ibrewise tensor
product is that the barycentric coordinates and the separations are independent.
Indeed, setting
\[
\bar{\bm q}=\frac1n(\bm q_1+\dotsb+\bm q_n),
\]
we f\/ind
\[
[\bar{\bm q}^\mu,\delta_{ij}\bm q^\nu]=0.
\]
This is true in a strong sense, so that \(\bar{\bm q}\) can be understood
to live in a dif\/ferent tensor
factor than the separations (up to an  isomorphism).
Then it is possible to set all the
\(\delta_{ij}\bm q^\nu\) to their minimum value, thus leaving a
function of \(\bar{\bm q}\) alone, with a hidden \(\sigma\) dependence
consisting of restriction to \(\Sigma_1\) (due to the fact that the Euclidean distance is not covariant under Lorentz boosts).

What we obtain in this way is a quantum analogue of the restriction
\[f(x_1,\dotsc,x_n)\restriction_{x_1=x_2=\dotsb=x}\] to the diagonal of a
function of many events \cite{piac_tesi,Bahns:2003vb}.

In the framework of Dubois--Violette  universal calculus \cite{Djemal:1995aaa},
we may def\/ine \(d\bm q=\bm q\otimes \bm I-\bm I\otimes\bm q=\bm q_2-\bm q_1\),
where, as everywhere in this section, we understand \(\otimes\) ``f\/ibrewise''
(or in the sense of \(Z\)-modules). A natural def\/inition of 4-volume
is then
\[
\bm V=d\bm q\wedge d\bm q\wedge d\bm q\wedge d\bm q=
\epsilon_{\mu\nu\rho\sigma}(\bm q_2-\bm q_1)^\mu(\bm q_3-\bm q_2)^\nu(\bm q_4-\bm q_3)^\rho(\bm q_5-\bm q_4)^\sigma
\]
(which is intuitively reminiscent of the volume of a hypercube),
and analogously for 3-volumes and area operators. The spectra of all these
operators can be computed; in particular the spectrum of
\(\bm V\) does not contain 0, and stays at a f\/inite
distance~\(\sim\lambda_P^4\) from it~\cite{4-volume}.

\subsection[Geometry of $\Sigma$; dilation covariance]{Geometry of $\boldsymbol{\Sigma}$; dilation covariance}

Given a set of pairwise commuting operators \(\bm Q^{\mu\nu}\) fulf\/illing
\(\bm Q^{\mu\nu}=-\bm Q^{\nu\mu}\), we may consider them as the entries
of an antisymmetric matrix with operator entries
\[
\bm Q=\begin{pmatrix}
        \bm Q^{00}&\bm Q^{01}&\bm Q^{02}&\bm Q^{03}\\
        \bm Q^{10}&\bm Q^{11}&\bm Q^{12}&\bm Q^{13}\\
        \bm Q^{20}&\bm Q^{21}&\bm Q^{22}&\bm Q^{23}\\
        \bm Q^{30}&\bm Q^{31}&\bm Q^{32}&\bm Q^{33}
      \end{pmatrix}
=     \begin{pmatrix}
        \bm 0&\bm e_1&\bm e_2&\bm e_3\\
       -\bm e_1&\bm 0&\bm m_3&-\bm m_2\\
       -\bm e_2&-\bm m_3&\bm 0&\bm m_1\\
       -\bm e_3&\bm m_2&-\bm m_1&\bm 0
      \end{pmatrix},
\]
where, following \cite{Doplicher:1994tu},
we introduced the notation\footnote{There
are slight dif\/ferences with formulas in~\cite{Doplicher:1994tu},
since here we def\/ine
$\vec{\bm e}$, $\vec{\bm m}$ with respect to indices in upper position.}
\(\bm e_j=\bm Q^{0j}\) and \(\bm m_i=\bm Q^{jk}\) (\((i,j,k)\) a cyclic
permutation of~\((1,2,3)\)). The (pseudo-)vectors $\vec{\bm e}$, $\vec{\bm m}$
are called the electric and magnetic parts of~\(\bm Q\), in {\itshape
formal} analogy with electromagnetism (no physical meaning attached to this
terminology). Note that no meaning is attached to the position of the
indices of $\vec{\bm e}$, $\vec{\bm m}$.

Correspondingly, we may represent the matrix \(\bm Q\) as a pair
\((\vec{\bm e},\vec{\bm m})\) of vectors with operator entries.
Now, given a matrix \(\sigma\in\Sigma\), we may
represent it as a pair \((\vec e,\vec m)\) of vectors with real entries;
we recall that, for the DFR model,
\(\text{spec}(\bm Q^{\mu\nu})=\{\sigma^{\mu\nu}:\sigma\in\Sigma\}\);
under the above cor\-respondence,
\(\text{spec}(\bm e_j)=\{\sigma^{0j}:\sigma\in\Sigma\}=
\{e_j:(\vec e,\vec m)\in\Sigma\}\) and
\(\text{spec}(\bm m_j)=\{m_j:(\vec e,\vec m)\in\Sigma\}\).

For the DFR model, \(\Sigma\) is the orbit of the matrix
\((\vec e_3,\vec e_3)\), where \(\vec e_3=(0,0,1)\) is the third vector
in the canonical basis of \(\mathbb R^3\).

The advantage of this notation is that the two invariants take a very simple
form:
\[
\bm Q_{\mu\nu}\bm Q^{\mu\nu}=2(|\vec{\bm m}|^2-|\vec{\bm e}|^2),\qquad
\bm Q_{\mu\nu}(\ast \bm Q)^{\mu\nu}=4(\vec{\bm m}\cdot\vec{\bm e}).
\]
It follows that
\[
\Sigma=\{(\vec e,\vec m): |\vec e|=|\vec m|, \vec e\cdot\vec m=\pm 1\}.
\]

We observed at the end of Section~\ref{sec:dfr_recovered} that the Planck
scale is given by the spectrum of the operator
\([\bm q^0,\bm q^1,\bm q^2,\bm q^3]\). There is however a natural
way
to construct a model with \([\bm q^0,\bm q^1,\bm q^2,\bm q^3]=0\), where
there is no characteristic length. This model is covariant
under dilations, too (S.~Doplicher, private conversation).

This can be obtained by replacing \(\Sigma\) with the orbit
\[
\Sigma_{\text{conf}}=\{(\vec e,\vec m):|\vec e|=|\vec m|,\vec e\perp\vec m\};
\]
for example, \((\vec e_1,\vec e_2)\in\Sigma_{\text{conf}}\), which corresponds
to the (non trivial) commutation relations
\[
[\bm X^0,\bm X^1]=[\bm X^1,\bm X^3]=i\bm I.
\]
Then by Schur's lemma \(\bm X^2\) and \(\bm X^0+\bm X^3\)
are multiples of the identity;
there are \(\infty^2\) equivalence classes of irreducible representations
corresponding to the given choice of \(\sigma=(\vec e_1,\vec e_2)\):
\[
(\bm X^0,\bm X^1,\bm X^2,\bm X^3)=
(\bm Q,\bm P,\alpha\bm I, \beta\bm I-\bm Q),\qquad \alpha,\beta\in\mathbb R.
\]
Note that this is dif\/ferent than in the \(\lambda_P\neq 0\) case, where to each \(\sigma\) there corresponds a unique class of equivalence classes. However,
essentially by the same methods of~\cite{Doplicher:1994tu}, it is possible to
 build up coordinates \(\bm q^\mu\) with commutators
\(\bm Q^{\mu\nu}=-i[\bm q^\mu,\bm q^\nu]\) fulf\/illing
\[
\bm Q^{\mu\nu}\bm Q_{\mu\nu}=\bm Q^{\mu\nu}(\ast\bm Q)_{\mu\nu}=0,
\]
and covariant under a unitary representation of the group generated
by Poincar\'e transformations and dilations. Uncertainty relations and
f\/ield theory will be analysed elsewhere.

\section{The ``canonical'' quantum spacetime}
\label{sec:CQST}

The so called ``canonical'' quantum spacetime is def\/ined by the relations
\begin{equation}
\label{eq:cqs_rel}
[\bm x^\mu,\bm x^\nu]=i\theta^{\mu\nu}\bm I.
\end{equation}
According to the discussion of Section~\ref{subsec:exist_coord},
the DFR argument grants the
existence of at least one regular representation if
there is a matrix \(\varLambda\) such that
\(\theta^{\mu\nu}={\varLambda^\mu}_{\mu'}{\varLambda^\nu}_{\nu'}S^{\mu'\nu'}\).

However,
if no such \(\varLambda\) is known to exist, everything is possible: there could be
other inequi\-va\-lent
regular representations, or no representations at all. Even if~\(\varLambda\)
exists, but is not invertible, there might be many other inequivalent
representations which cannot be obtained in this way.
In these cases, no general solution is known, and
the representation theory of the given relations
must be discussed case by case.

We consider two classes of examples in this section.
In the next section, we discuss the representation theory of
models where the time coordinate
commutes with all the space coordinates (\(\theta^{0j}=0\)), and we show
that the existence of representations depends on the particular choice
of~\(\theta\) (and as a byproduct, we describe the method for
classifying them all in the good cases).

In the rest of the  section
we will discuss the case where \(\theta\) is a DFR matrix, in which case
the representations always exist.

We will shortly comment on the conceptual implications
of the breakdown of isotropy in the f\/lat spacetime at Planck scale,
and on the lack of motivations.

\subsection{Time/space commutative models and representations}
\label{subsec:sad_fate}

We consider the most general time/space commutative models where
\(\theta^{0j}=0\), so that
in parti\-cu\-lar \([\bm x^0,\bm x^j]=0\).

Space/time commutative models have no direct physical motivations; they
enjoy some fortune because they apparently remove an obstruction to the
development of a unitary non local perturbative \(S\)-matrix. We shall
later see that these obstructions only are due to an improper treatment
of time ordering, the consequences of which
are hidden by time/space commutativity.

For some \(a,b,c \in\mathbb R\), we have
\[
\theta=\begin{pmatrix}
0&0&0&0\\
0&0&-a&-b\\
0&a&0&-c\\
0&b&c&0
\end{pmatrix},
\]
which means
\[
[\bm x^0,\bm x^k]=0,\qquad
[\bm x^1,\bm x^2]=-ia\bm I,\qquad
[\bm x^1,\bm x^3]=-ib\bm I,\qquad
[\bm x^2,\bm x^3]=-ic\bm I,
\]
and where at least one of $a$, $b$, $c$ is not zero. We f\/irst seek
for irreducible representations.

By Schur's lemma, \(\bm x^0=q\bm I\) for some \(q\in\mathbb R\).
Now, according to the desired relations, the operator
\(\bm T=c\bm x^1 -b\bm x^2+a\bm x^3\) should fulf\/il
\([\bm T,\bm x^\mu]=0\) and thus, again by Schur's lemma, \(\bm T\)~should be a multiple
of the identity. We assume, say, that \(a\neq 0\);
otherwise one may reason analogously (or permute the indices). Then it should be
\[
\bm x^3=\frac1a(b\bm x^2-c\bm x^1)+q'\bm I
\]
for some \(q'\in\mathbb R\). In other words, the representation is irreducible
if and only if  $\bm x^1$, $\bm x^2$ are irreducible; but in view of this
and their commutation relations,
by von Neumann uniqueness we must have
$\bm x^1=\bm P$, $\bm x^2=a\bm Q$ (up to equivalence)
where \([\bm P,\bm Q]=-i\bm I\).

To sum up, the most general irreducible representation should have the form
\[
\bm x^0=q\bm I,\qquad\bm x^1=\bm P,\quad \bm x^2=a\bm Q,\qquad
\bm x^3=\frac ba \bm P-\frac ca \bm Q+q'\bm I,
\]
for some \(q,q'\in\mathbb R\). But in this way, we f\/ind
\[
[\bm x^1,\bm x^3]=i\frac ca\bm I,\qquad
[\bm x^2,\bm x^3]=ib\bm I;
\]
which are the desired relations if and only if $c=-ab$, $c=-b$
(with solutions $a=1$, $c=-b\in\mathbb R$ or $a\in\mathbb R$, $c=b=0$); this gives
conditions for the existence of irreducible representations in the case \(a\neq 0\). In other words,
the relations~\eqref{eq:cqs_rel} admit representations only for some choices of~\(\theta\). For the good choices, all irreducible representations are obtained
in the above way.

Note that in the bad case where no representation exists, there is no Weyl
quantisation, and there is no corresponding Weyl product. Of course the usual
formula for the Weyl product still is meaningful and may well be taken as
a standalone def\/inition; but the resulting $*$-algebra would not admit
 $C^*$-completions.

On the other side, let us now assume that we are in the good case,
and that representations exist.
Hence we classif\/ied them all.
Note however that, if we took any of such irreducible representations
as our choice of the coordinates, we would f\/ind
\(e^{ik_\mu\bm x^\mu}=e^{ik^0 q}e^{-i\vec k\cdot \vec{\bm x}}\) for the Weyl
operators, which would give a very strange quantisation prescription, whose
ef\/fect on the time variable would be evaluation at the f\/ixed value
\(x^0=q\):
\[
\int_{\mathbb R^4} dk\; e^{ik_\mu\bm x^\mu}\check f(k)=
\frac1{(2\pi)^3}
\int_{\mathbb R^3} d\vec x\int_{\mathbb R^3} d\vec k
f(q,\vec x)e^{-i\vec k\cdot(\vec{\bm x}-\vec x)}.
\]
Apart from describing a very strange classical limit with \ldots\ constant time
\(x^0=q\), it would be impossible to def\/ine an associated
twisted product (which requires injectivity of the quantisation, not to be ill
posed).

Hence we have no rights to make an arbitrary choice, and we must take into
account all irreducible representations at once.
By direct integral techniques\footnote{Note that
\(\int_{\mathbb R}^\oplus dq\;\mathbb C=L^2(\mathbb R)\),
and \(\int_{\mathbb R}^\oplus dq\;q\cdot=\bm Q\),
the Schr\"odinger position.}, we get the universal representation
(in the case \(a=b=-c=1\), say)
\[
\bm \xi^0=\bm Q_2,\qquad
\bm \xi^1=\bm P_1,\qquad
\bm \xi^2=\bm Q_1,\qquad
\bm \xi^3=\bm P_1+\bm Q_1+\bm Q_3,
\]
where $\bm Q_1$, $\bm Q_2$, $\bm Q_3$,  are the Schr\"odinger position operators
for a particle in the 3-space,
and \(\bm P_1\) is the Schr\"odinger momentum fulf\/illing
\([\bm P_1,\bm Q_k]=-i\delta_{1k}\bm I\).
By construction (or a direct check), they
fulf\/il
\[
[\bm\xi^\mu,\bm\xi^\nu]=i\theta^{\mu\nu}\bm I
\]
as required. The Weyl quantisation is injective, the twisted product is well
def\/ined, and there is a unique $C^*$-completion of the resulting algebra, which
is \(\mathcal C_0(\mathbb R^2)\otimes\mathcal K\).

\subsection{``Canonical'' quantum spacetime and DFR model}

From now on, we stick to the case of DFR matrices,
where \(\theta=\lambda_P^2\sigma\)
for \(\sigma\in\Sigma\). In this case the
``canonical'' quantum spacetime
is nothing but one single f\/ibre over the DFR $C^*$-bundle
\(\mathcal E=\mathcal C_0(\Sigma,\mathcal K)\), precisely the f\/ibre
over the chosen \(\sigma\). Abstractly, the
f\/ibre is the same for all \(\sigma\)s,
hence the choice of a particular \(\sigma\) only entails a particular
choice of the Weyl quantisation prescription, and the corresponding
(reduced) twisted product.
Let us write
\[
\bm q^\mu_{(\sigma)}=\lambda_P{\varLambda^\mu}_\nu\bm X^\mu,
\]
where the \(\bm X^\mu\)'s are def\/ined in Section~\ref{subsec:exist_coord}
and \(\varLambda\) is any\footnote{The matrix \(\varLambda\) is def\/ined
up to elements of the stabiliser of \(S\) in \(\mathscr L\); however
\(\bm q^\mu_{(\sigma)}\) does not depend on this choice; it only depends
on \(\sigma\).} Lorentz matrix fulf\/illing
\(\sigma^{\mu\nu}={\varLambda^\mu}_{\mu'}{\varLambda^\nu}_{\nu'}S^{\mu'\nu'}\).
Then by construction
\begin{equation}\label{eq:sigma_rels}
[\bm q^\mu_{(\sigma)},\bm q^\nu_{(\sigma)}]=i\lambda_P^2\sigma^{\mu\nu}\bm I,
\end{equation}
and the quantisation over \(\sigma\) is def\/ined by
\[
f(\bm q_{(\sigma)})=\int dk\;\check f(k)e^{ik_\mu\bm q^\mu_{(\sigma)}};
\]
it maps admissible symbols into compact operators on \(L^2(\mathbb R)\), and
fulf\/ils
\[
(f\star_\sigma g)(\bm q_{(\sigma)})=f(\bm q_{(\sigma)})g(\bm q_{(\sigma)}).
\]

In mathematics, ``canonical'' is used to indicate something which is
independent from arbitrary choices. Standing the arbitrariness of the choice
of a particular \(\sigma\), the terminology ``canonical quantum spacetime''
is totally unjustif\/ied.

As it stands, the spacetime quantisation associated with the reduction to a
particular irreducible representation is clearly non covariant, and there is not
much to say.

However, it is possible to use the DFR bundle of algebras to def\/ine
a form covariant model, where however the relativity of observers
is broken, and Wigner's approach to quantum symmetries is dismissed.
It amounts to let  \(\theta\) transform as a tensor, and to attach to each
reference frame \(\mathcal O'\) its own \(\theta'\), compatibly with Lorentz
transformations of reference frames.
This model had some fortune in the literature, and can be shown to be
equivalent to yet another approach named ``twisted covariance'', which we shall
discuss in Section~\ref{subsec:tw_cov}.

The basic idea is to f\/ix a specif\/ic \(\theta\) with respect to a specif\/ic
frame, and claim that the observer~\(\mathscr O\) in that particular
frame ``sees'' the commutation relations \eqref{eq:sigma_rels}. Let us call
it the privileged observer.

Then an observer \(\mathscr O'\) in a frame related to the privileged
frame by a  Poincar\'e transformation \((\varLambda,a)\) will correspondingly
``see'' the commutation relations
\[
[\bm q^\mu_{(\sigma')},\bm q^\nu_{(\sigma')}]=i\lambda_P^2{\sigma'}^{\mu\nu}\bm I,
\]
The resulting formalism is then form-covariant; yet it is possible to classify
the observers in an absolute way, according to the relations they ``see''.
While in the DFR model all the observers (in the sense of special relativity)
are equivalent, here there are inf\/initely many equivalence classes of
observers, labeled by \(\Sigma\); two observers are equivalent if and only
if they are connected by a Poincar\'e transformation \((\varLambda,a)\)
such that \(\varLambda\) leaves \(S\) unchanged (=\(\varLambda\)
is in the stabiliser of \(S\) in~\(\mathscr L\)).

\subsection{``Canonical quantum spacetime'' and localisation states}
\label{subsec:cqst_inv_red}
The situation of the end of the preceding section may be described
as regarding the DFR bundle of algebras as a collection of algebras
labeled by \(\sigma\), together with a groupoid of automorphisms
connecting pairs of algebras in that collection; the global algebraic
structure carried by the f\/ibrewise product of sections is dismissed.

An equivalent way of describing it is to retain the global algebraic
structure, while restricting instead the class of admissible localisation
states \cite{Piacitelli:2009fa}. The best way to understand this argument is
to think that there are the Gods,
who can see the whole structure of the spacetime; and the poor human
beings (the observers described by the theory), with limited
capability of understanding the Universe.

Indeed, observers can test the algebra with the states at their disposal.
Let us assume that, by a decision of the Gods,
the privileged observer \(\mathscr O\) only may test, in his own
reference frame, the full DFR algebra with states such that each
\(\bm Q^{\mu\nu}\) is completely determined and has expectation~\(\sigma^{\mu\nu}\):
\[
\omega(\bm Q^{\mu\nu})=\sigma^{\mu\nu},\qquad
\Delta_{\omega}(\bm Q^{\mu\nu})=0.
\]
We denote by
\(\mathscr S_{\mathscr O}\) this class of states.
By def\/inition, for any state \(\omega\in\mathscr S_{\mathscr O}\)
there is a state
\(w_{\omega}\) on \(\mathcal K\) such that
\[
\omega(f(\bm Q;\bm q))=w_\omega(f(\sigma;\bm q_{(\sigma)})),
\]
and all states in \(\mathscr S_{\mathscr O}\) arise in this way. It follows that
on one side
\[
\omega(f(\bm Q;\bm q)g(\bm Q;\bm q))=w_\omega(f(\sigma;\bm q_{(\sigma)})
g(\sigma;\bm q_{(\sigma)})),\qquad \omega\in\mathscr S_{\mathscr O};
\]
on the other,
\[
\omega((f\star g)(\bm Q;\bm q))=w_\omega((f(\sigma;\cdot)\star_\sigma
g(\sigma;\cdot))(\bm q_{(\sigma)})),\qquad\omega\in\mathscr S_{\mathscr O}.
\]
In other words, being enabled to test the geometry only by means of states
in \(\mathscr S_{\mathscr O}\), the privileged observer will not recognise
the full algebraic structure available to the Gods, and only will f\/ind
the ``canonical quantum spacetime'' over \(\sigma\);
his symbol algebra
will be \(\mathcal E_{\sigma}\), his twisted pro\-duct~\(\star_\sigma\),
and so on and so forth.

Let us now give a look at the situation of
the primed observer \(\mathscr O'\), connected to \(\mathscr O\)
by the Poincar\'e transformation \((\varLambda,a)\). The states available to
her are just those in the pull-back
\[
\mathscr S_{\mathscr O'}=
\{\omega(\bm U(\varLambda,a)^{-1}\cdot\bm U(\varLambda,a)):\omega\in
\mathscr S_{\mathscr O},(\varLambda,a)\in\mathscr P\}
\]
of \(\mathscr S_{\mathscr O}\).
They are precisely the states such that each \(\bm Q^{\mu\nu}\) is
completely determined and has expectation \({\sigma'}^{\mu\nu}\).
By repeating the discussion, his symbol algebra
will be~\(\mathcal E_{\sigma'}\), his twisted pro\-duct~\(\star_{\sigma'}\),
and so on and so forth.

Hence we are in the following situation: there is a perfectly covariant
model, the DFR model, where the equivalence of observers is fully enforced.
Fixing
a \(\theta=\lambda_P^2\sigma\)
in a particular frame is equivalent to put a non invariant constraint on
that model, which amounts to dismiss a huge class
of otherwise admissible localisation states by means of a non invariant
selection criterion. The natural question is then:
{\scshape{Why should we dismiss all those states?}}
This question was raised in~\cite{Piacitelli:2009fa,Piacitelli:2009tb}.

\subsection{Twisted covariance}
\label{subsec:tw_cov}
In this section, we discuss
an apparently dif\/ferent
approach to covariance, based on quantum deformations of the Lorentz
group (in the spirit of quantum groups)~\cite{Chaichian:2004za,Wess:2003da}.

Following \cite{Piacitelli:2009fa,Piacitelli:2009tb},
we will convince ourselves
that this formalism is equivalent in spirit to work with the full DFR model,
if we agree to dismiss a large classes of otherwise admissible localisation
states (in the sense described in Section~\ref{subsec:cqst_inv_red});
``in spirit''  meaning: up to the choice between
Weyl quantisation and algebraic relations.

Again, we will take a bottom-up approach, and describe twisted covariance by
adding further degrees of structure step by step, when necessary. This will
help us to keep track of the various assumptions, and of the nature of the
mathematical concepts.

\subsubsection*{Twists}
We will start by considering a map \(f\otimes g\mapsto f\star g\) of the form
\[
(f\star g)(x)=\int da\; db\; K_2(x,a,b)f(a)g(b),\qquad x\in \mathbb R^4.
\]
The only specif\/ic initial assumption is that manipulations like{\samepage
\[
\int da\int db=\int db\int da,\qquad  \lim\int=\int\lim
\]
are allowed (possibly in a weak sense). Functions will be assumed smooth at
wish.}

Under the above assumptions, for some \(\xi_2\) such that \(\xi_2(x,x)=x\),
we may def\/ine
\begin{gather*}
F_2(x,y,a,b)=K_2(\xi_2(x,y),a,b),\\
({\mathcal F_2} f\otimes g)(x,y)=\int da\;db\;F_2(x,y,a,b)f(a)g(b).
\end{gather*}
With
\[
m_2: \ f\otimes g\mapsto fg,\qquad \tilde m_2: \ f\otimes g\mapsto f\star g,
\]
it follows that
\[
\tilde m_2= m_2\circ {\mathcal F}_2.
\]
Note that we did not use associativity of \(\star\), nor invertibility
of \(\mathcal F_2\).

\subsubsection*{Finite twisted covariance}

The only additional assumption now is invertibility of
\(\mathcal F_2\).

With \(\mathscr L\) the Lorentz group, we def\/ine the actions
of \(\varLambda\in\mathscr L\)
\[
(\gamma(\varLambda)f)(x)=f(\varLambda^{-1}x),\qquad
\gamma_2(\varLambda)=\gamma(\varLambda)\otimes\gamma(\varLambda);
\]
they are intertwined by the usual product:
\begin{equation}
\label{eq:cov}
m_2\circ\gamma_2(\varLambda)=\gamma(\varLambda)\circ m_2.
\end{equation}

The twisted action of \(\mathscr L\) is def\/ined as
\[
\tilde\gamma_2(\varLambda)=
{\mathcal F}_2^{-1}\circ \gamma_2(\varLambda)\circ {\mathcal F}_2;
\]
it is conceived so that, by construction,
\begin{equation}
\label{eq:tw_cov}
\tilde m_2\circ\tilde\gamma_2(\varLambda)=\gamma(\varLambda)\circ\tilde m_2,
\end{equation}
which may be regarded as a deformation of \eqref{eq:cov}.
Note that in \eqref{eq:tw_cov} the action \(\gamma(\varLambda)\) on functions
of one variable is unchanged. Equation \eqref{eq:cov}
expresses usual covariance of the pointwise product;
\eqref{eq:tw_cov} is called twisted covariance of the twisted product.

We will need the following commutation relation:
let \({\mathcal F}^{(\varLambda)}_2\) be def\/ined by
\begin{equation}\label{eq:comm_F}
\gamma_2(\varLambda)\circ {\mathcal F}_2=
{\mathcal F}_2^{(\varLambda)}\circ\gamma_2(\varLambda);
\end{equation}
then of course
\[
({\mathcal F}_2^{(\varLambda)}f\otimes g)(x,y)=\int da\;db\;
F_2(\varLambda x,\varLambda y,\varLambda a,\varLambda b)
f(a)g(b).
\]

\subsubsection*{Inf\/initesimal twisted covariance}
Here too, we only assume invertibility of \(\mathcal F_2\).
Let us def\/ine \(X^\mu\) by setting \((X^\mu f)(x)=x^\mu f(x)\);
with \(\omega\in \text{Lie}(\mathscr L)\), we have the action
\[
\omega\triangleright{} ={\omega^\mu}_\nu X^\nu\partial_\mu,
\]
so that
\[
\gamma(e^{t\omega})f=f-t\omega\triangleright f+o(t).
\]
Let \(\Delta\) be the coproduct for the usual (undeformed) coalgebra structure
of the the universal enveloping
Lie algebra \(U(\text{Lie}(\mathscr L))\) of the Lie algebra
of \(\text{Lie}(\mathscr L)\) of the Lorentz group.
With
\[
\Delta(\omega)=\omega\otimes 1+1\otimes \omega,
\]
we have the action
\[
\Delta(\omega)\triangleright{}=(\omega\triangleright{})\otimes I+
I\otimes(\omega\triangleright{})
\]
so that
\[
\gamma_2(e^{t\omega})f\otimes g=f\otimes g-
t\Delta(\omega)\triangleright f\otimes g+o(t).
\]
Moreover,
\[
\tilde\gamma_2(e^{t\omega})f\otimes g=f\otimes g-
t\widetilde{\Delta(\omega)\triangleright{}}f\otimes g+o(t),
\]
where
\[
\widetilde{\Delta(\omega)\triangleright{}}=
{\mathcal F}_2^{-1}\circ(\Delta(\omega)\triangleright{})\circ {\mathcal F}_2.
\]
Note that the above def\/ines a {\itshape twisted action} of the {\itshape
undeformed} coproduct.

By \eqref{eq:comm_F},
\[
\tilde\gamma_2(\varLambda)={\mathcal F}_2^{-1}\circ {\mathcal F}^{(\varLambda)}\circ
\gamma_2(\varLambda).
\]
Hence
\begin{gather*}
\widetilde{\Delta(\omega)\triangleright{}}=
\left.\frac{d}{dt}
\left({\mathcal F}^{-1}\circ {\mathcal F}^{(e^{-t\omega})}\circ
\gamma_2(e^{-t\omega})\right)\right|_{t=0}
= (\Delta(\omega)\triangleright{})+
{\mathcal F}^{-1}\left.\frac{d}{dt}{\mathcal F}^{(e^{-t\omega})}\right|_{t=0}.
\end{gather*}

\subsubsection*{Weak coassociativity}
Besides invertibility of \(\mathcal F_2\),
now we also assume that \(\star\) is associative,
namely \((f\star g)\star h=f\star(g\star h)\). With \(K_3\) def\/ined by
\[
(f\star g\star h)(x)=\int da\;db\;dc\;
K_3(x;a,b,c)f(a)g(b)h(c),
\]
associativity implies that
\[
K_3(x;a,b,c)=\int dy\;K_2(x;y,c)K_2(y;a,b)=\int dy\;K_2(x;a,y)K_2(y;b,c)
\]
{\itshape within integrals}.
We may now reproduce all the steps: for some \(\xi_3\) such that
\(\xi_3(x,x,x)=x\), def\/ine
\[
(\mathcal F_3f\otimes g\otimes h)(x,y,z)=\int da\;db\;dc\;
K_3(\xi_3(x,y,z);a,b,c)f(a)g(b)h(c);
\]
with \(m_3f\otimes g\otimes h=fgh\), set
\[
\tilde m_3(f\otimes g\otimes h)=
f\star g\star h=m_3\circ \mathcal F_3 f\otimes g\otimes h;
\]
if \(\mathcal F_3\) is invertible, set
\begin{gather*}
\gamma_3(\varLambda)=
\gamma(\varLambda)\otimes\gamma(\varLambda)\otimes\gamma(\varLambda),\\
\tilde\gamma_3=\mathcal F_3^{-1}\circ\gamma_3\circ\mathcal F_3=
\mathcal F_3^{-1}\circ\mathcal F_3^{(\varLambda)}\circ\gamma_3
\end{gather*}
it follows that
\[
\tilde m_3\circ\tilde\gamma_3(\varLambda)=\gamma(\varLambda)\circ\tilde m_3.
\]
As for inf\/initesimal transformations,
\[
\tilde\gamma_3(e^{t\omega})f\otimes g\otimes h=
(\text{id}_3+\widetilde{\Delta_3(\omega)\triangleright{}}
+o(t))f\otimes g\otimes h,
\]
where
\begin{gather*}
\widetilde{\Delta_3(\omega)\triangleright{}}= \mathcal F_3^{-1}
\circ\Delta_3(\omega)\circ\mathcal F_3
= \Delta_3(\omega)\triangleright{}+
\left.\mathcal F_3\circ \frac{d}{dt}\mathcal F_3^{(e^{-t\omega})}\right|_{t=0}.
\end{gather*}
Above \(\Delta_3(\omega)\) is obtain by iteration of the undeformed,
coassociative coproduct:
\[
\Delta_3(\omega)=(\Delta(\omega)\otimes 1)\circ\Delta(\omega)=
(1\otimes\Delta(\omega))\circ\Delta(\omega).
\]
Now, if we choose \(\xi\) symmetric, namely
\[
\xi(x_1,x_2)=\xi(x_2,x_1),
\]
then we have weak twisted coassociativity, namely coassociativity
of the twisted action:
\[
\widetilde{\Delta_3(\omega)\triangleright{}}=
((\widetilde{\Delta(\omega)\triangleright{}})\otimes I)
\circ\widetilde{\Delta(\omega)\triangleright{}}=
(I\otimes(\widetilde{\Delta(\omega)\triangleright{}}))
\circ\widetilde{\Delta(\omega)\triangleright{}}.
\]

\subsubsection*{Strict twisted covariance}
The above steps where performed under rather mild assumptions. Maybe
the most restrictive assumption is the invertibility of the
twist operators; in the cases where they turn out to be invertible,
it only
requires some care with the analytic aspects to obtain weakly
coassociative twisted covariance for a suf\/f\/iciently regular model
based on the Weyl quantisation of some set of relations.

However, in the above setting the Hopf algebra \(U(\text{Lie}(\mathscr L))\)
associated with the Lie group
\(\mathscr L\) remains in the background:
the coproduct is the usual one, twists only af\/fect its action.

We now make the f\/inal step: the additional bit of structure
is to assume that \(\mathcal F_2=F_2\triangleright\) for some
(invertible) \(F_2\in U(\text{Lie}(\mathscr L))\) which fulf\/ils
some requirements (enforcing twisted coassociativity). In that case we can drop
the symbol \(\triangleright\) and def\/ine a new coproduct
\[
\tilde\Delta(\cdot)=F_2^{-1}\Delta(\cdot)F_2.
\]

Hence we have the following situation.
Weakly twisted Poincar\'e covariance is possible
both for the ``canonical'' quantum spacetime \cite{Piacitelli:2009tb,Piacitelli:2009fa} (see also next paragraph) and the
\(\kappa\)-Minkowski spacetime~\cite{Dabrowski:2009mw,Dabrowski:2010im}.
In the f\/irst case, we can make the last step to strict twisted covariance~\mbox{\cite{Chaichian:2004za,Wess:2003da}}.  In the second case, strict twisted
Poincar\'e covariance is not possible in the case of \(\kappa\)-Minkowski
spacetime because of the obstructions to the existence of a suitable
\(F_2\) (although this obstruction can be circumvented by adding the dilation
group, see~\cite{2010arXiv1005.4429B}).

\subsubsection*{Canonical quantum spacetime and twisted covariance}

Let \(\sigma\in\Sigma\) and $\bm q^\mu_{(\sigma)}$, $\star_\sigma$ be the
corresponding irreducible coordinates and star product.

We may restrict ourselves to the Schwartz functions, which are dense
in the algebra of symbols, and allow for most manipulations with
integrals and limits.

We recall that, like the usual pointwise product becomes convolution
under Fourier transform (where \(\widehat{fg}=\hat f\times\hat g\)),
its deformation \(\star_\sigma\) is related with
the corresponding deformation \(\tilde\times_\sigma\)
of the convolution product again
by Fourier theory:
\[
\widehat {f\star_\sigma g}=\hat f\tilde\times_\sigma\hat g,
\]
where
\[
(\hat f\tilde\times\hat g)(k)=\int dk\;f(h)g(k-h)e^{\frac i2
h_\mu\sigma^{\mu\nu} k_\nu}.
\]
Let us introduce the notation
\[
c(\hat f\otimes \hat g)=\hat f\times \hat g
\]
for the usual (undeformed) convolution; then we have
\[
c_\sigma(\hat f\otimes \hat g)=\hat f\tilde\times_\sigma \hat g=
(c\circ \mathcal T_\sigma)f\otimes g,
\]
where \(\mathcal T_\sigma\) is simply the multiplication by
\(e^{(i/2)\sigma^{\mu\nu}h_\nu\otimes k_\mu}\), and is evidently invertible.
Hence an invertible \(\mathcal F_\sigma\) exists, and we f\/ind
\[
f\star_\sigma g=m\circ\mathcal F_\sigma(f\otimes g).
\]
Relation \eqref{eq:comm_F} now reads
\[
\gamma_2(\varLambda)\circ {\mathcal F}_\sigma=
{\mathcal F}_{\sigma'}\circ\gamma_2(\varLambda),
\]
where primes indicate usual Lorentz actions on scalars, 4-vectors,
and functions. It follows that
\[
m_\sigma\circ\tilde\gamma^{(2)}(\varLambda)(f\otimes g)=
m\circ\mathcal F_\sigma\circ
\mathcal F^{-1}_\sigma(\mathcal F_\sigma^{-1}f\otimes g)'=
f'\star_{\sigma'}g'=(f\star_\sigma g)'.
\]

Hence the usual form-covariance (where \(\sigma\) is treated as a
tensor and the action of Lorentz transformations is the usual one, in the
framework described in Section~\ref{subsec:cqst_inv_red})
is perfectly equivalent~--~as a formalism~--~to weakly coassociative
twisted covariance, where
\begin{itemize}\itemsep=0pt
\item \(\sigma\) is treated as a constant,
\item the Lorentz action on functions of one event is unmodif\/ied, and
\item the Lorentz action on functions of two or more events is twisted.
\end{itemize}
In turn, we have seen that usual form-covariance applied to ``canonical''
quantum spacetime is equivalent to deal with the full DFR model,
up to dismiss a huge class of otherwise admissible localisation states (Section~\ref{subsec:cqst_inv_red}).

\section{Quantum f\/ield theory}
\label{sec:qft_qst}

\subsection{Local quantum f\/ields}
It is wise to shortly revise the fundamental concepts underlying
relativistic quantum f\/ield theory on classical
spacetime\footnote{Sometimes this is referred to as ``commutative quantum
f\/ield theory'' (CQFT), as opposed to NCQFT; this is a nonsensical terminology,
see also footnote~\ref{fn:NC} at page~\pageref{fn:NC}.}, for the purpose
of identifying the concepts which will have to be modif\/ied, and the motivations
for such modif\/ications.

To formulate a relativistic quantum physics, it is necessary to f\/ind a way
to establish a quantum version of Einstein locality (or causality,
they are synonyms). The most
natural way is to rely on the usual notion of statistical independence for
quantum observables, namely commutativity. If two observables $\bm A$, $\bm B$
are localised
at some events $x$, $y$ respectively, they
must fulf\/il \([\bm A,\bm B]=0\) whenever \(x-y\) is spacelike\footnote{We
take the signature \((+---)\) for the Lorentz metric; hence \(x\) is spacelike
if \(x_\mu x^\mu<0\).}.

The theory is said local if the above condition is fulf\/illed, and any
possible observable is either localised, or in the algebra generated by the
localised observables.

To sum up, in order to describe a relativistic theory,
two new concepts (axioms?)
are to be injected into quantum physics: 1)~it must be
meaningful to ask whether a certain observable is localised in any region
of spacetime (ideally even at a  sharp point) or not, and 2)~Einstein causality must hold
in the form of commutativity at
spacelike distances (we conf\/ine ourselves to observable f\/ields).

The other basic ingredient is a unitary representation \(\bm U\)
of the Poincar\'e group \(\mathscr P\), implementing a symmetry of the system
with the natural geometric interpretation;
namely for any observable \(\bm A\) localised at \(x\),
\(\bm U(\varLambda,a)\bm A\bm U(\varLambda,a)^{-1}\) represents
the same  experimental procedure, either performed
\begin{itemize}\itemsep=0pt
\item  in a dif\/ferent, equivalent laboratory  as seen by the initial observer
in her reference frame (active point of view), or
\item in the original event, but seen by a dif\/ferent observer
from his reference frame (passive point of view).
\end{itemize}
Note that, for the active point of view to be
meaningful, the localisation of the device must belong to the specif\/ications
of the experimental setup associated with the observable.

The action of translations def\/ine the total energy-momentum \(\bm P^\mu\)
of the theory by \(\bm U(I,a)=e^{ia_\mu \bm P^\mu}\). In non relativistic
quantum physics it is necessary to postulate that the energy is lower bounded
(otherwise an inf\/inite amount of energy could be extracted from the system).
The relativistic version is that the joint spectral resolution
\(\bm P^\mu\ket p=p^\mu\ket p\) of \(\bm P\) must fulf\/il the condition
\(p_\mu p^\mu\geqslant 0\). In the absence of spontaneously broken symmetries,
we also require uniqueness of the lower energy state \(\ket{0}\), called
the vacuum.

Let \(\bm A\) be an observable localised at the event \(x_0\); then for any
other event \(x\) we may def\/ine a new observable \(\bm\phi(x)\) by setting
\[
\bm\phi(x)=\bm U(I,x_0-x)\bm A\bm U(I,x_0-x)^{-1},
\]
where we take the active point of view. In other words, \(\bm\phi(x)\)
describes the observable obtained by displacing the experimental setup of
\(\bm\phi(x_0)=\bm A\) from its original event \(x_0\) into the new event~\(x\).
We are then naturally led to consider operator-valued ``functions'' of the
spacetime. The active point of view forces us to require consistence with Einstein locality:
\[
[\bm\phi(x),\bm\phi(y)]=0,\qquad \text{\(x-y\) spacelike}.
\]
Moreover, the above must be compatible with Poincar\'e symmetry:
\[
\bm U(\varLambda,a)\bm\phi(x)\bm U(\varLambda,a)^{-1}=
\bm\phi(\varLambda^{-1}(x-a)).
\]
A f\/ield fulf\/illing the above two conditions (Einstein locality and Poincar\'e
covariance) is called a relativistic quantum f\/ield. If in a theory
there are many independent observable f\/ields\footnote{Although for
general reasons the theory might encompass unobservable f\/ields
with dif\/ferent commutation relations (related with global gauge symmetries)
at this level this is irrelevant for our discussion.} \(\bm\phi_j\),
they must be relatively local:
\[
[\bm\phi_j(x),\bm\phi_k(y)]=0,\qquad \text{\(x-y\) spacelike}.
\]

For several reasons (see \cite{corinaldesi} for a review) the above picture
is too optimistic: quantum f\/ields are too singular, and cannot be treated as
ordinary functions; sharply localised f\/ields ``\(\bm\phi(x)\)'' only are
meaningful within integrals (namely: as distributions). While this fact
should not necessarily be seen as disturbing
(pointwise localised instruments would have been an idealisation in any case),
it might be regarded as the very f\/irst
manifestation that something might go wrong in the inf\/initely small.

In standard textbooks, this is usually taken care of at the technical level
by assuming that quantum f\/ields are elements of some class of operator valued
distributions, def\/ined on a suitable class of regular test functions. However
in this way we miss a point which is very important to us: that in this game
test functions do not play the r\^ole of elements of an algebra of functions;
they should instead be thought of as linear
functionals on that algebra!

So, let us elaborate this idea of ``smearing''. As we have seen,
already on classical spacetime, the localisation of an observable
must be made ``fuzzy'' by choosing a probability
density \(\rho(x)dx\), and smearing
the f\/ield over that density:
\[
\bm\phi(\rho)=\int\bm\phi(x)\rho(x)dx.
\]
Since f\/ields are too singular, these probability
densities must be suf\/f\/iciently regular
(we usually take inf\/initely dif\/ferentiable functions), and vanish at inf\/inity
suf\/f\/iciently fast (usually one takes functions with compact support or faster than inverse polynomials). In particular, a probability measure of the form
\(\delta(x-a)dx\) (sharp localisation at \(a\in\mathbb R^4\)) is not available.

Since any complex function with the same properties (smoothness, fast decay)
can be written as the linear combination of at most four regular
probability densities, it is
natural to dismiss the requirements of positivity
and normalisation, and extend the f\/ields to generic test functions by
linearity: f\/ields are usually def\/ined as linear maps
\(\ell\mapsto\bm\phi(\ell)\) from test functions
(= linear functionals, not necessarily positive and/or normalised)
to operators.

The notation \(\ell\) for a test
function is unconventional; the reason is that we wish to reserve the symbol
\(f\) for an element of the localisation algebra
\(\mathcal C_0(\mathbb R^4\)).
If \(\rho\) is a probability density, then
\[
f\mapsto\int f(x)\rho(x)dx
\]
is a well def\/ined state on the localisation algebra
\(\mathcal C_0(\mathbb R^4)\). Moreover, any test function
\(\ell\) def\/ines a continuous linear functional
\[
f\mapsto\int f(x)\ell(x)dx.
\]
Hence, quantum f\/ields \(\bm\phi(\cdot)\) should be thought of
as maps from the (suf\/f\/iciently regular)
states of the localisation algebra to the operators
on some Hilbert space; extended by linearity to the regular linear functionals.

Then an expression like \(\int P(\bm(\phi(x))\ell(x)dx\), where \(P\) is some
polynomial (e.g.\ a Wick polynomial), is to be interpreted in terms
of pointwise (local) products of f\/ields, smeared with a~suf\/f\/iciently regular linear functional \(\ell\) on the localisation algebra.

\subsection{DFR quantisation of local quantum free f\/ields}

Let \(\bm F\) be a continuous function of \(\mathbb R^4\), vanishing at inf\/inity,
and  taking values in the operators on some Hilbert space (or in some
$C^*$-algebra \(\mathfrak F\)). Then we may set
\[
\bm F(\bm q)=\int dk\;e^{ik_\mu\bm q^\mu}\otimes \check{\bm F}(k),
\]
where
\[
\check{\bm F(k)}=\frac{1}{(2\pi)^4}\int dx\;\bm F(x)e^{-ik_\mu x^\mu},
\]
is well def\/ined (provided $\bm F$, $\check{\bm F}$ are in \(L^1\)).

We may apply the above to a function of the form \(\bm F(x)=f(x)\bm A\),
where \(\bm A\) is some f\/ixed operator (or element of \(\mathfrak F\)),
and \(f\) is a complex continuous function, and we get \(\bm F(\bm q)=
f(\bm q)\otimes \bm A\).
If we consider another function of the form \(\bm G(x)=g(x)\bm B\), then we get
\(\bm G(\bm q)=g(\bm q)\otimes \bm B\).
Their product has two sources of non-commutativity:
the original one, due to the fact that the functions take values in a non-commutative algebra; and the new one, due to the spacetime quantisation.
In other words, already before spacetime quantisation the pointwise
product was non-commutative:
\((\bm{FG})(x)=\bm F(x)\bm G(x)\neq \bm G(x)\bm F(x)=(\bm{GF})(x)\),
in general. Now spacetime quantisation in a sense
``increases the non-commutativity''.

The most general function \(\bm F\) as above can be approximated as
\begin{equation}\label{eq:tensor_approx}
\bm F(x)\approx \sum_jf_j(x)\bm A_j,
\end{equation}
so that
\begin{equation}\label{eq:q_tensor_approx}
\bm F(\bm q)\approx \sum_jf_j(\bm q)\otimes \bm A_j;
\end{equation}
note that above \(\bm F(\bm q)\) is obtained by applying the DFR quantisation
of ordinary symbols in the f\/irst tensor factor only,
{\itshape leaving the second tensor factor unmodified.}

The exact mathematical meaning of \eqref{eq:tensor_approx} is that
there is a canonical isomorphism between
 \(\mathcal C_0(\mathbb R^4,\mathfrak F)\) and
\(\mathcal C_0(\mathbb R^4)\otimes\mathfrak F\), sending the function
\(x\mapsto f(x)A\) into the element \(f\otimes A\). The meaning of~\eqref{eq:q_tensor_approx} is that the quantisation of \(\bm F(x)\) is obtained
by combining the above mentioned isomorphism with the ordinary quantisation
on the f\/irst tensor factor only.

{\sloppy The above comments embody the statement that the non-commutative
replacement of \(\mathcal C_0(\mathbb R^4,\mathfrak F)\) is
\(\mathcal E\otimes\mathfrak F\). But we have much more, we have a consistent
quantisation recipe for the \(\mathfrak F\)-valued functions.

}

We may of course def\/ine a symbolic calculus, but, precisely as in the
case of ordinary symbols, we must allow for more general symbols, namely
\(\mathfrak  F\)-valued functions \(\bm F=\bm F(\sigma;x)\)
of \(\Sigma\times\mathbb R^4\), and def\/ine their quantisation
\(\bm F(\bm Q;\bm q)\) consistently. This allows to def\/ine a twisted product
by
\[
\bm F(\bm Q;\bm q)\bm G(\bm Q;\bm q)=(\bm F\star \bm G)(\bm Q;\bm q),
\]
which again is f\/ibrewise:
\[
(\bm F\star \bm G)(\sigma;x)=
\bm F(\sigma;\cdot)\star_\sigma \bm F(\sigma;\cdot).
\]

Now, a local free f\/ield \(\bm\phi\) is not continuous (it's a distribution),
nor it takes values in a $C^*$-algebra  (f\/ields are unbounded operators,
in general).
However, precisely as in the classical case, they can be formally
treated as ordinary  functions to some extent
(see any book on Wightman theory). Hence, following \cite{Doplicher:1994tu},
we may give the formal recipe
\begin{equation}
\bm\phi(\bm q)=\int dk\;e^{ik_\mu \bm q^\mu}\otimes\check{\bm\phi}(k)
\end{equation}
for the quantisation of a given local free quantum f\/ield.

Note that, precisely in the same way as the label ``\(x\)'' of a local quantum
f\/ield \(\bm\phi(x)\) is not an observable,
here the ``\(\bm q\)'' of  \(\bm\phi(\bm q)\) is not an observable!

Let \(\omega\) be a localisation state on the quantum spacetime
localisation algebra. We can apply it to the f\/irst tensor factor,
and obtain
\[
\bm\phi(\omega)=\int dk\;(\omega\otimes\text{id})(e^{ik_\mu \bm q^\mu}\otimes\check{\bm\phi}(k))=
\int dx\;\bm\phi(x)\rho_\omega(x),
\]
where
\[
\rho_\omega(x)=\frac{1}{(2\pi)^4}\int dk\;\omega(e^{ik_\mu\bm q^\mu})
e^{-ik_\mu\bm x^\mu};
\]
namely the initial local f\/ield \(\bm\phi(x)\) evaluated on a probability
density on the classical spacetime. This gives us the def\/inition
of ``suf\/f\/iciently regular state \(\omega\)'', as a state \(\omega\)
such that \(\rho_\omega\) is an admissible test function for the initial f\/ield.

The question as to why we restricted ourselves to free f\/ields has a
simple and neat answer: no interacting f\/ield is known in four
dimensions.

\subsection{Perturbation theory -- the Dyson series}
Local perturbation theory of local quantum f\/ields on classical spacetime
(in its more sophisticated version, namely renormalisation
theory) has been very ef\/fective in giving extremely accurate predictions
with wonderful experimental validation. Quite surprisingly, it works
even if the interaction is not a ``small'' perturbation of the free dynamics
in any reasonable sense.

As a f\/irst attempt to develop quantum f\/ield theory on quantum spacetime, we may
expect that non local Planck scale corrections are ``small'' enough not to
destroy the good behaviour of the local perturbation series. On the contrary,
the fuzziness of quantum spacetime should have an intrinsic regularising
ef\/fect on the ultraviolet (i.e.\ small scale) divergences.

We begin with an important remark.
Let \(\mathscr H(\phi,\partial_\mu\phi)\) be the usual (Wick ordered)
Hamiltonian density of the free scalar (Klein--Gordon) f\/ield \(\bm\phi\), with
\(\bm H_0=\int_{x_0=t} d\vec x
\mathscr H(\bm\phi(x),\partial_\mu\bm\phi(x))\)
(as an operator on the Fock space; it does not depend on \(t\)); then
it was found in \cite{Doplicher:1994tu} that
\[
\int_{q_0=t} d^3q\;\mathscr H(\bm \phi(\bm q),\partial_\mu\bm
\phi(\bm q))=\bm H_0\qquad \text{(as a constant function of \(\sigma\))}.
\]
The exact meaning of \(\int_{q_0=t} d^3q\) is explained in Section~\ref{subsec:working} (see in particular equation~\eqref{eq:pos_exp}).

The above remark suggests that the free theory remains consistent after
spacetime quantisation. This is a rewarding conf\/irmation, but not really a
surprise. Indeed spacetime quantisation is a purely kinematical, fully covariant
procedure, and the local free f\/ield is well def\/ined and covariant as well.

The next step is to def\/ine, for each \(t\),
\[
\bm H_I(\bm Q;t)=\int_{q_0=t} d^3q\;\wick{\bm\phi(\bm q)^n},
\]
which may be thought of as  a non constant function of \(\sigma\)
(see Section~\ref{subsec:working}). In symbolic language, we may rewrite
the above as
\[
\bm H_I(\sigma;t)=\int dx\;\delta(x^0-t)
\wick{(\bm\phi\star_\sigma\dotsm\star_\sigma\bm\phi)(x)}.
\]

Up to now, everything was perfectly satisfactory. However now we have the
problem of the \(\sigma\) dependence of \(\bm H_I\). At some point of the story, it must be
integrated out. The reason is that this is a model for particle scattering.
Particles scattered far away are
free and thus do not undergo high energy processes which could
excite the quantum geometric background. They are to be described by
the dear old free f\/ields (as in- and out-f\/ields).

There are many inequivalent possibilities. But they all are
af\/fected by the same problem: {\itshape there is no Lorentz invariant
measure on \(\Sigma\)}. It's just a fact of life\footnote{Technically it is a consequence of \(\mathscr L^+\) being not amenable.}. Hence whatever measure
is chosen over \(\Sigma\), this operations will break Lorentz covariance.
We will come back to this problem later on.

The choice proposed in \cite{Doplicher:1994tu} is to take the rotation invariant measure
\(d\sigma\) on \(\Sigma_1\subset\Sigma\),
where~\(\Sigma_1\) was described at the end
of Section~\ref{subsec:working}. This choice is for the largest possible
geometric symmetry.
We get
\[
\bm H_I(t)=
\int_{\Sigma_1} d\sigma\int dx\;\delta(x^0-t)
\wick{\bm\phi\star_\sigma\dotsm\star_\sigma\bm\phi}(x)
\]
as a non local replacement of the usual local perturbation
of the Hamiltonian of the free f\/ield, in the interaction
picture. This interaction is covariant under translations and space
rotations, but not under Lorentz boosts.

The opposite choice is to take for the measure on \(\Sigma\) the Dirac measure
concentrated on some special choice of \(\sigma\), which gives the
same result as if we would have restricted ourselves from the start to the
``canonical quantum spacetime'' corresponding to that choice of \(\sigma\).

Note that, for any of the above choices, there is a suitable kernel \(G_t\)
such that
\begin{equation}\label{eq:nonloc_int}
\bm H_I(t)=
\int_{\mathbb R^{4n}} da_1\dotsm da_n\;G_t(a_1,\dotsc,a_n)
\bm\phi(a_1)\dotsm\bm\phi(a_n).
\end{equation}

The scattering matrix (\(S\)-matrix) is def\/ined as the formal solution
\[
S=U(\infty,-\infty)
\]
of
\[
\frac{d}{dt}\bm U(t,s)=i\bm H_I(t)\bm U(t,s)
\]
fulf\/illing \(\bm U(t,t)=\bm I\). It can be
described in terms of Dyson's non-commutative
modif\/ication of Picard's method:
\[
S=\sum_{n=0}^\infty \frac1{n!}\int dt_1\dotsm dt_n
T[\bm H_I(t_1) , \dots,\bm H_I(t_n)];
\]
the time ordered product is def\/ined as{\samepage
\[
T[ \bm H_I(t_1) , \dots,\bm H_I(t_n)]=\bm H_I(t_{\pi(1)})\dotsm\bm H_I(t_{\pi(n)}),
\]
where \(\pi\) is the permutation of \((1,2,\dots,n)\) such that
\(t_{\pi(1)}>t_{\pi(2)}>\dots >t_{\pi(n)}\).}

The regularising ef\/fect of non locality on the Dyson series for this
ansatz has not yet been fully investigated. However there are interesting
partial results; in particular Bahns~\cite{Bahns:2004zb} found an ultraviolet f\/inite
$S$-matrix in a variant of this model for the \(\wick{\bm\phi^{\star 3}}\)
interaction, where the \(\sigma\) variable is
integrated out independently at any vertex of the resulting diagrams.

Other proposals, which go beyond the scope of this review, are possible
for the generalisation of Wick product on quantum spacetime. Indeed,
on classical spacetime \(fg\) can be looked at in at least two ways:
as a  pointwise
product \(f\otimes g\mapsto fg\) in the algebra of functions, or
as a~limiting procedure \((fg)(x)=\lim\limits_{y\rightarrow x}f(x)g(y)\).
These two equivalent procedures have inequivalent generalisations to quantum
spacetime:
the f\/irst becomes \(f\otimes g\mapsto f\star g\), while the second
can be given a meaning by sending the dif\/ferences \(\bm q_j-\bm q_k\)
to their minimum, compatibly with positivity (i.e.\ uncertainty relations),
as described in Section \ref{subsec:quantum_geom}.
The f\/irst one is precisely the one used in~\cite{Doplicher:1994tu}, which we already have
seen. The second is investigated in~\cite{Bahns:2003vb} and used to def\/ine another form of quantum Wick product;
it results in a complete ultraviolet regularisation of the $S$-matrix.
A third proposal is to give standalone
def\/inition of ``local'' subtractions of divergences for the product
\(\bm\phi^{\star n}\); this gives yet another def\/inition of quantum
Wick pro\-duct~\cite{Bahns:2004fc}, whose behaviour in the Dyson series has
not yet been investigated.

Not only there are many inequivalent generalisations of the Wick product.
There also are inequivalent generalisations of the perturbative series.
The Yang--Feldman equation, which is equivalent to the Dyson series
on the classical spacetime, provides a dif\/ferent evolution series~\cite{Bahns:2002vm}.

In all the approach described above, interactions sooner or later destroy
Lorentz covariance. A striking example of this
situation is provided by the approach based on the
Yang--Feldman equation, which seems to be covariant at all steps; yet
in the end it requires a \(\sigma\)-dependent mass renormalisation
\cite{Bahns:2004fc}.

Since every more or less naive generalisation of local interactions
seems to lead to a conf\/lict with Lorentz covariance, it is
reasonable to conjecture that the crucial point on which we should
concentrate our ef\/forts is to understand the fate of
locality beyond non-commutativity.

\subsection{Formal unitarity is not violated!}
\label{subsec:un_viol}

In \cite{Gomis:2000zz},  unitarity violations of the \(S\)-matrix were
found for the \(\phi^{\star 4}\) theory,
in the form of a failure of the optical theorem for a particular graph
(the ``f\/ish'').

However, it is clear that, if the interaction Hamiltonian is formally
selfadjoint, the \(S\)-matrix is formally unitary, and no formal violations
of unitarity can be expected. So, what went wrong?

There are two main sources of ambiguity: the Euclidean methods and
the Moyal expansion. We f\/irst discuss the latter. To this end we consider
the f\/irst order contribution to the two points \(\tau\)-function,
arising upon insertion of the Dyson expansion
into the Gell-Mann--Low formula\footnote{Here we omit the (inf\/inite)
normalisation \(\tfrac1{\bra0S\ket0}\); for an explicit proof that
the cancellation of vacuum-vacuum components carries over also in the non local framework, see \cite{Piacitelli:2004rm}.}:
\begin{equation}\label{eq:2order_GML}
-\frac12\int dt\;\bra0 T[ \bm\phi(x) , \bm\phi(y) , \bm H_I(t) ]\ket 0,
\end{equation}
where the time ordering refers to the variables $x^0$, $y^0$, $t$.

For the sake of comparison we f\/irst recall a basic trick of the local theory,
in the case of an interaction term of the form \(H_I(t)=\int d^4a\;\delta(t-a^0)
\wick{\bm\phi^n(a)}\). Plugging it into the above formula, and carrying f\/irst
the integration over the variable \(t\) we get
\[
-\frac12\int_{\mathbb R^4} da
\bra0T[ \bm\phi(x) , \bm\phi(y) ,
\wick{\bm\phi^n(a)} ]\ket0,
\]
where now the time ordering refers to the variables $x^0$, $y^0$, $a^0$.
Inadvertently, the time ordering has been shifted inside the integral
which appear in the def\/inition of \(\bm H_I(t)\).
The reason why this is possible is that the time parameter def\/ining the
interaction term is precisely the same as the time variable
the Wick monomials are evaluated at; hence integration and time ordering
commute.

The same trick cannot be done if instead we plug a non local
interaction term of the
form~\eqref{eq:nonloc_int} into~\eqref{eq:2order_GML}, which gives
\[
-\frac12\int dt\;\bra0T\left[\phi(x),\phi(y), \int da_1\dotsm da_n
G_t(a_1,\dots,a_n)\wick{\bm\phi(a_1)\dotsm\bm\phi(a_n)} \right]\ket0,
\]
where the time ordering is relative to the variables $x^0$, $y^0$, $t$ and
{\itshape has no relations at all with the variables \(a_j^0\).}
This comment was already available in~\cite{Doplicher:1994tu}, but was
rediscovered much later: the correct time ordered prescription was then given
the name of ``Interaction Point Time Ordering Prescription'' (IPTOP), which
is a somewhat unfortunate terminology, since a basic feature on
non commutative spacetimes is that the concept of point is not available
any more.

Indeed, it was shown in~\cite{Bahns:2002vm} that, with the proper
treatment of time ordering, the optical theorem for the ``f\/ish'' graph
holds true, and there is no violation of unitarity even if time
and space do not commute. Indeed, this is only one single example;
formal unitarity was already implied by the choice of using the Dyson series
with the correct time ordering prescription.

It has been observed that, if one takes a
time/space commutative ``canonical quantum spacetime'', than the violations
of unitarity disappear, and this was taken as an indication that time/space
non-commutativity was responsible for the unitarity violations.
As we have seen, this is not the case. Formal unitarity violations were due
to an improper treatment of the time ordering; the only ef\/fect of
time/space commutativity is to cancel the ef\/fects of that error.

We observed in Section~\ref{subsec:misuses} that, in a sense, the Moyal
expansion is non-commutative but local; non-locality being taken care of
by analyticity. If one manipulates the Moyal expansion and formally treats the
twist as if it were a true dif\/ferential operator, then all the f\/ields (external
and internal vertices of Feynman diagrams) would be {\itshape apparently}
evaluated at the same time and  we would be naturally
led to bring the time ordering inside the integral, thus taking
the wrong time ordering prescription.

Of course, this is not a proof that there exists a unitary theory;
indeed, this would require having completed the renormalisation programme in
the Minkowskian setting. On the contrary, this programme is quite
underdeveloped, since it must be formulated in terms of operator f\/ields, which
is a formidable task already in the commutative case (this is precisely the
reason of the fortune of Euclidean measure-theoretic methods on
classical spacetime).

\subsection{Perturbation theory and diagrams}

The perturbative terms  in the Gell-Mann--Low formula can be treated in a very
economical way by means of Feynman diagrams, namely simple drawings
representing complicate integrals, encoded in rules for drawing the diagrams.
Unfortunately the situation in the
non-commutative case is quite confusing.

The f\/irst set of rules for diagrams appeared in a paper by Thomas Filk
\cite{Filk:1996dm}. Filk's approach was in the spirit of a standalone $S$-matrix
theory, as described e.g.\ in~\cite{tHooft:1973pz}\footnote{It is well known that
this approach was
f\/irst sponsored by Werner Heisenberg; according to a private
conversation reported by Rudolph Haag, however, Heisenberg then
dismissed this view.}. So, his starting point was
to consider the classical action, keep the quadratic (i.e. free) term
unchanged (according to DFR analysis),
and replace the pointwise product in the interaction term
with the twisted product.

Unfortunately, the classical action arises
in the Gell-Mann--Low formula precisely due to the trick of bringing the time
ordering inside the integral (see any standard textbook on QFT); which is not
allowed  in the case of time/space non-commutativity.
The Filk rules, then, can not be derived from the Hamiltonian evolution proposed
in~\cite{Doplicher:1994tu}.
This is the f\/irst manifestation of a basic fact of life: it is
not obvious that perturbative methods which are equivalent in the commutative
case have equivalent non-commutative generalisations; usually they don't.
In this case, the lack of unitarity could be regarded as a good reason to
dismiss this approach.

This also explains why Filk rules turn out to fulf\/il unitarity in the case
of time/space commutativity. Motivated by \cite{Bahns:2002vm}, the rules for
the DFR Hamiltonian approach were developed in~\cite{Liao:2002xc,Liao:2002pj}
(see also the very clear~\cite{Bahns:2004mm}). In view of a remark of~\cite{Denk:2003jj}, the situation was further clarif\/ied in~\cite{Piacitelli:2004rm}, to which we may refer for more details.

Additional sets of inequivalent rules for diagrams can be obtained
in the Euclidean framework, discussed in the next section.

\subsection{Euclidean methods and UV/IR mixing}
There is a well known connection  between local QFT and classical statistical
mechanics. Indeed, due to translation covariance and the spectrum condition,
vacuum expectations
\[
\mathscr W(x_1-x_2,x_2-x_3,\dots, x_{n-1}-x_n)=
\bra 0\bm\phi(x_1)\bm\phi(x_2)\dotsm\bm\phi(x_n)\ket 0
\]
of local product of f\/ields (Wightman functions) have analytic continuations to
imaginary time (Wick rotation), and the resulting functions \(S_n(\eta_1,\dots,\eta_{n-1})\) (the Schwinger functions) are totally symmetric; here
\(\eta_j^0=i(x_j-x_{j+1})^0\).
This suggests to think of
them as of correlation functions of generalised Brownian motions, were the
space of ``paths'' is a suitable space of distributions on \(\mathbb R^4\),
called Euclidean f\/ields (or sometimes Euclidean f\/ield conf\/igurations).
The probability measure tolling these ``paths'' is expressed in terms of
the classical Euclidean action, which arises from the combination of the
Wick rotation  and of locality in time (the shift of time ordering inside the
integral, discussed in Section~\ref{subsec:un_viol}).
This allows for using methods from stochastic theory in the
computation of Feynman diagrams; the idea is to f\/inally
switch back to real time to obtain physically
observable quantities (inverse Wick rotation).
This approach is called ``Euclidean'' because
the Lorentzian square of a 4-vector is positive def\/inite at imaginary
times, if the signature \(-+++\) is taken.

Note that the Euclidean formulation is possible only at the level of
expectation values (Wightman functions);
no Euclidean operator f\/ield theory can be directly obtained by a Wick
rotation, because of the sad fate of the time
evolution operator \(e^{it\bm H}\), when computed at imaginary times.

Wightman functions enjoy a property called ``Wightman positivity'', which
is equivalent to the positivity of transition probabilities. Thanks to locality
and covariance, this property has a Euclidean counterpart,
called ``ref\/lection positivity'',
or ``Osterwalder--Schrader'' positivity; it is necessary for a
family of would-be Schwinger functions to be actually connected to
some physical expectation values in Minkowski space (see e.g.\ the nice and
readable textbook~\cite{Roepstorff:1994ga}). Note however that the
correspondence between Wightman and OS axioms is not one-to-one;
OS positivity is a
joint consequence of Wightman positivity, locality and covariance.

It is evident that there are many potential sources of trouble in trying to
extend this method to the non commutative case:
\begin{itemize}\itemsep=0pt
\item[a)] the original argument showing that the analytic continuation reaches
imaginary times still may be valid in the case of time space commutativity,
since it is based on the spectral condition (see d) here below);
\item[b)] since Lorentz covariance is broken,
Schwinger functions cannot be expected to be symmetric, so their
interpretation as correlation functions is lost, and the stochastic
interpretation (if any) cannot be expected to be obtained by
a simple deformation of the local partition function (which would give
symmetric Schwinger functions);
\item[c)] no replacement for ref\/lection positivity is available so far
(note that it should reproduce usual ref\/lection positivity
in the large scale limit).
\end{itemize}
The above remarks are general;
if in addition time does not commute with space,
\begin{itemize}\itemsep=0pt
\item[d)] twists
blow up exponentially at imaginary times and destroy the analyticity
argument for the Wick rotation based on the spectrum condition
(see however \cite{Bahns:2009iq});
\item[e)] there is really no reason to expect any
r\^ole for the classical Euclidean action, because of the time ordering
issue.
\end{itemize}

Euclidean theory can (and is) of course be studied as a standalone
$S$-matrix theory, but it is not clear which could be its physical content;
without any guidance from f\/irst principles or experiments,
there is little hope to guess
appropriate Euclidean Feynman rules, giving a physical theory on Minkowski
space up to some Wick rotation. Note that
it also might be sensible to expect such a theory to have a local large
scale limit.

Indeed, there are good
indications that the current Euclidean approach is not related with
the Minkowski formulation \cite{Bahns:2009iq}.

Since the standalone approach is rather fashionable, quite
often authors do not state explicitly which is their perturbative setting.
The joint consequences of Euclidean methods and
exotic time ordering prescription are then quite dif\/f\/icult to
disentangle. Writing diagrams in terms of the classical action would
mean to take an exotic time ordering prescription on Minkowski space,
if the Euclidean theory could be obtained by a naive Wick rotation (it can not).

An example where such a disentanglement would be particularly useful is in
the issue of IR/UV mixing, namely the conjecture that
infrared and ultraviolet divergences
(usually de\-coupled in local QFT) get coupled in the non commutative case.
This was f\/irst observed in the Euclidean setting in \cite{Minwalla:1999px},
where
the following intuitive explanation is given:
``Roughly, very small pulses
instantaneously spread out very far upon interacting.
In this manner very high energy
processes have important long distance consequences.''
Were this true, then we could expect single
particle collisions at Planck energy to have large detectable ef\/fects at
astronomical distances, which would be quite surprising. The IR/UV mixing
is now the strongest obstruction to renormalisation in the Euclidean setting;
as we said, in the Minkowskian setting renormalisation
is way underdeveloped, and it is not even crystal clear how to formulate it.
As of today we do not know if
IR/UV mixing (or similar features/pathologies) does arise as well in the
Minkowskian setting (this claim could become outdated in the near future).

\subsection*{Acknowledgements}
I am indebted with Dorothea Bahns, Ludwik Dabrowski, Sergio Doplicher,
Klaus Fredenhagen and Giorgio Immirzi for
stimulating conversations, and for comments on a preliminary version. Dorothea
Bahns drove my attention on reference~\cite{cuntz-weyl-algebra}.

\LastPageEnding

\end{document}